\documentclass{article}

\usepackage{PRIMEarxiv}

\usepackage[utf8]{inputenc}
\usepackage[T1]{fontenc}
\usepackage[hidelinks]{hyperref}
\usepackage{url}
\usepackage{booktabs}
\usepackage{nth}
\usepackage{natbib}
\usepackage{fancyhdr}
\usepackage{graphicx}
\usepackage{placeins}

\title{Bay Assessment Model: A Python-Based Salinity Projection Tool for Florida Bay, USA}

\author{ Erik Stabenau\thanks{Corresponding author. ORCID
    \href{https://orcid.org/0000-0002-6574-9317}{0000-0002-6574-9317}} \\
  The Everglades Foundation \\
  18001 Old Cutler Road, Palmetto Bay, FL 33157, USA \\
  \texttt{EStabenau@EvergladesFoundation.org} \\
  \And
  Joseph Park\thanks{ORCID
    \href{https://orcid.org/0000-0001-5411-1409}{0000-0001-5411-1409}} \\
  Sugihara Lab, Scripps Institution of Oceanography \\
  9500 Gilman Drive, La Jolla, CA 92093, USA \\
  Biological Nonlinear Dynamics Data Science Unit \\
  Okinawa Institute of Science and Technology Graduate University \\
  1919-1 Tancha, Onna, Okinawa 904-0495, Japan \\
  \texttt{JosephPark@ieee.org} \\
}

\begin{document}
\maketitle

\begin{abstract}
Extreme salinity events in Florida Bay are a central concern of the
Comprehensive Everglades Restoration Program (CERP). A large-scale seagrass
die-off in the late 1980s, driven by hypersaline conditions following reduced
freshwater delivery, made reliable salinity prediction essential to restoration
planning. The Bay Assessment Model (BAM) is a hydrologic model that simulates
salinity across 54 idealized basins representing Florida Bay. It is
mass-conservative apart from where required for model stability. Interbasin
fluxes are computed from hydraulic gradients across shoals, and each basin
receives direct rainfall and evapotranspiration forcing. Along the shoreward
boundary BAM is driven by observed water levels from the Everglades Depth
Estimation Network (EDEN), or by output from upstream restoration planning
models. Along the marine margins, tidal boundary conditions from NOAA
subordinate station harmonic constituents are superimposed on a regional mean
sea level anomaly. BAM is implemented entirely in Python and is open source.
Over the extended 1999--2026 record the domain-wide mean water level bias is
$-$0.024 m with a mean RMSE of 0.092 m, and the domain-wide mean salinity bias
is 0.00 ppt with a mean RMSE of 6.21 ppt. Model bias varies systematically with
hydrologic regime, increasing in magnitude from drought to wet conditions and
reversing sign at one central bay basin, constraining the interpretation of
anomaly-based metrics. Applying uniform offsets at the Everglades shoreward
boundary yields salinity responses of 2--4 ppt in the northeastern bay,
attenuating toward the marine margins; a fuller treatment of restoration and
sea level scenarios is reported separately. BAM provides a transparent,
efficient, and community-accessible tool for evaluating the salinity
consequences of Everglades restoration actions in Florida Bay.
\end{abstract}


\section{Introduction}

Early in the \nth{20} Century, wetland reclamation efforts in 
South Florida, USA, were developed to remove vast quantities of 
freshwater from the state. The primary method for this was impoundment 
of waters behind dams and channelization of natural spillways to 
enhance de-watering after storms and throughout the wet season. 
The ecological impacts of these actions were first discovered along 
the coasts, where freshwater wells became salty and freshwater rivers 
became ingress points for saltwater. Along the coast, the rapid changes 
in salinity brought on by canal discharges and extended periods without 
freshwater brought on ecological stress. Eventually, Florida Bay at the 
southern end of the peninsula, became hypersaline. It experienced a 
large-scale seagrass and fish die-off event in the late 1980s \citep{Robblee1991}. 
This event called attention to the environmental issues plaguing the state 
and was a driving force behind Everglades restoration efforts. In the 
intervening years, restoration efforts have had to balance the benefits 
to the marsh ecosystems upstream and marine systems along the coast, 
largely in absence of appropriate salinity prediction tools. The Florida 
Bay Assessment Model (BAM) fills this need.

Prior efforts to estimate Florida Bay salinity span a range of 
approaches from empirical statistical models to full three-dimensional 
hydrodynamic simulations \citep{Marshall2008}. Empirical and regression-based tools, 
such as those described by \citet{Marshall2011}, derive salinity 
estimates directly from observed stage and flow relationships and 
are computationally efficient, but their lack of an explicit 
physical process representation limits their ability to project 
salinity response under boundary conditions outside the historical 
calibration record -- precisely the conditions of interest for 
restoration planning. Fully resolved hydrodynamic models, such as 
the coupled surface-water/groundwater simulation of 
\citet{Langevin2004}, offer greater physical fidelity but at a 
computational and data cost that has limited their routine use in 
iterative restoration scenario evaluation. BAM's immediate 
predecessor, the FATHOM model \citep{Cosby2010}, occupies an 
intermediate position with a basin-and-shoal representation similar 
to BAM's own, but relies on a coarser and shorter observational 
record and a compiled, non-public codebase. BAM is intended to 
close this gap by providing a physically based, computationally efficient, and 
fully transparent tool suited to the routine, iterative salinity 
assessment that restoration planning requires.

BAM is a 'basins' hydrological model of Florida Bay. It is
mass-conservative apart from where required for model stability, and
explicitly designed to predict salinities
based on water levels in 54 idealized basins representing Florida Bay.
In Florida Bay, the natural basins are separated and connected by shoals, 
thereby the model conforms to a linked-node network hierarchy with basins 
as nodes and shoals as links. Interbasin fluxes are driven by hydraulic 
gradients developed across the shoals in response to water level 
elevations and are modeled with depth integrated velocity based on 
frictional flow across shallow banks.

Each basin is forced with rainfall and evaporation. Basins on the Gulf of Mexico 
and Atlantic Ocean boundaries are also forced across the appropriate shoals with 
sea levels consisting of tidal variations and sea level changes. Coastal basins 
along the Everglades are forced with water levels determined from the Everglades 
Depth Estimation Network (EDEN) \citep{Telis2014} with the shoal properties 
(length, width, depth) calibrated to match aggregate runoff from the FATHOM 
model \citep{Cosby2010}.

BAM can be considered a derivative work from FATHOM since it employs the same 
basic physical and domain representations, however, it is significantly different 
in several aspects. First, BAM uses contemporary, high-quality, high-temporal 
density observational data to drive model inputs and boundaries. These environmental 
forcings leverage the wealth of meteorological and hydrographic observations 
available from the Marine Monitoring Network administered by Everglades National 
Park. Tidal data are computed from local NOAA subordinate tide stations within 
Florida Bay containing all astronomical forcings including intrannual variations. 
The physical description of shoals linking the Everglades and Atlantic Ocean 
to Florida Bay have been updated to more accurately reflect existing flow paths. 
The reliance on high-quality, nearly continuous observational data to drive 
model inputs is a particular strength of BAM.

Second, BAM is pure Python. While this imposes an operational constraint since 
model run-times are slower than would be the case with a compiled binary image, 
it affords several advantages. The use 
of object oriented code throughout facilitates clarity and enhances capacity for future
collaboration. The code is human readable using modern programming standards and naming
conventions. Using Python native data containers and dictionaries simplifies and further
clarifies the processing of data and objects. Finally, given the ubiquitous nature of Python,
the model can run on virtually any modern computing platform or operating system. Finally,
by sticking to Python there is no need for commercial compilers or other licensing that
could limit the use of the model.

Third, BAM outputs are not limited to specific time intervals, the output 
interval is user-controllable. Fourth, all BAM inputs/outputs are contained 
in ASCII comma delimited files imposing a uniform and accessible standard for 
model I/O enabling the evaluation of alternative scenarios in a straightforward 
manner. Fifth, the graphical user interface is designed to allow interaction 
and querying of model results and parameters. Each basin and shoal in the 
model is accessible with a mouse click on the map, or from a list box. Integrated 
plotting facilities allow the comparison of model outputs from different runs. 
Sixth, the GUI is not required. BAM can be run in text-mode inside a terminal 
and is therefore easily controlled in batch mode by shell or other scripts. 
Seventh, all model controls are exposed through the command line interface which 
applies whether the GUI is used or not.

Lastly, BAM is open-source. This fosters transparency, allows community 
development, and empowers the individual to adjust and change any aspect 
of the model to suit their needs or conceptualizations.

The remainder of this paper is organized as follows. 
Section~\ref{sec:description} describes the model domain, physical 
representation, and boundary forcing. Section~\ref{sec:data} 
documents the observational data and input files used for this 
effort. Section~\ref{sec:calibration} briefly summarizes the 
calibration approach. Section~\ref{sec:performance} evaluates model 
performance against the full 1999--2026 observational record for 
both water level and salinity. Section~\ref{sec:scenarios} 
illustrates model sensitivity to changes in Everglades freshwater 
delivery and mean sea level through a series of prescribed boundary 
condition scenarios. Section~\ref{sec:discussion} discusses model 
strengths, limitations, and appropriate applications, and 
Section~\ref{sec:conclusions} summarizes the principal findings. 
The BAM source code and operational manual are freely available; 
see Appendix A for access details.

\section{Model Description and Conceptual Framework}
\label{sec:description}

Florida Bay is most easily described as a shallow estuary consisting of 
semi-isolated basins, each sufficiently shallow (<2 m) and broad to be considered 
well mixed for modeling purposes. These basins are connected by extremely 
shallow shoals, with flow between basins driven by hydrologic gradients. The 
rate of water movement is primarily modeled on Manning's flow over these 
shoals, providing for conservation of mass. This simplified physical basis 
requires several assumptions that should be considered when interpreting results. 
Fundamentally, on each time step, the stage, volume, and mixing are treated as 
instantaneous. The model time step defaults to 6 min. with typical output 
water level, basin volume, and salinity averaged at the daily, weekly, or 
monthly level.

Bathymetry of the basins and shoals are binned at 0.305 m (1 ft.). This single 
layer model references water levels against a zero shoal depth. Changes to the 
shoal structure over time (e.g. due to erosion or sediment transfer) are ignored. 
Further, although each basin is unique in terms of depth and area, evaporation 
is applied uniformly across all but four basins in the model domain and precipitation is estimated 
from a limited set of daily rainfall data.

\subsection{Domain Representation}

\begin{figure}[htbp]
  \centering
  \includegraphics[width=\textwidth]{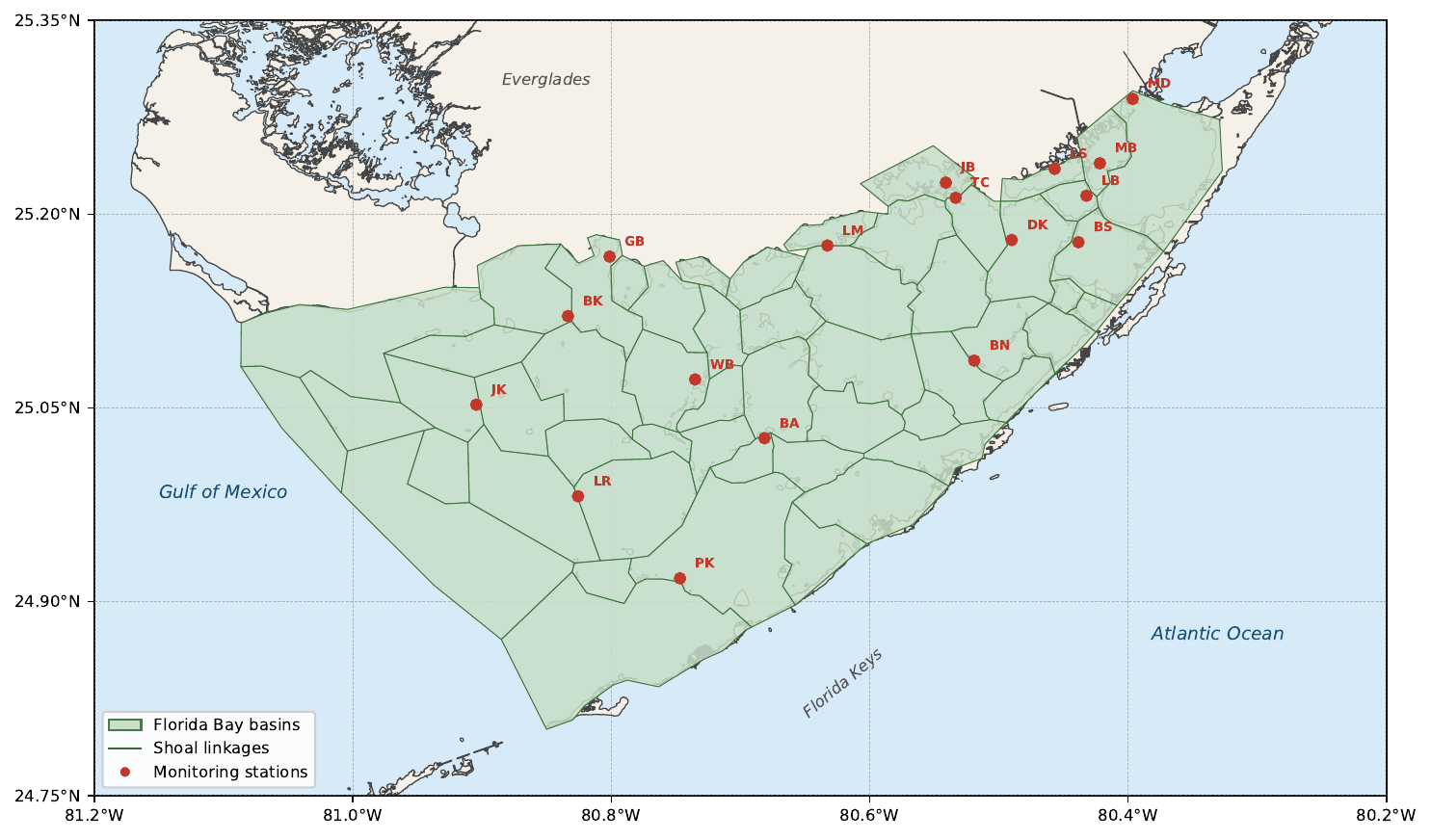}
  \caption{BAM model domain showing the 54 idealized basins representing 
  Florida Bay (green polygons), 410 shoal linkages (green lines), and the 
  17 Everglades National Park Marine Monitoring Network hydrographic stations 
  (red circles). Basin boundaries follow the mangrove fringes, buttonwood 
  embankments, and carbonate mud banks that naturally partition the bay into 
  semi-isolated hydrologic units. The shoreward boundary where Everglades 
  freshwater enters the domain lies along the northern margin; the Gulf of 
  Mexico and Atlantic Ocean marine boundaries lie to the west and southeast 
  respectively.}
  \label{fig:domain}
\end{figure}

For the purposes of the model, and much like its actual structure, 
Florida Bay is decomposed into 54 idealized basins whose boundaries 
are drawn along the mangrove fringes, buttonwood embankments, and 
carbonate mud banks that naturally partition the bay into semi-isolated 
hydrologic units (Figure~\ref{fig:domain}). Each basin is characterized 
by a depth-area relationship binned into up to 10 depth classes at 
0.305 m (1 ft.) increments, derived from available bathymetric surveys. 
Basin area and volume are computed from these depth classes and updated 
at each model time step as water levels change, providing a 
geometrically consistent representation of storage across the domain.

Basins are linked by 410 shoals, each representing a flow pathway 
across a bank, through a channel, or along a mangrove margin connecting 
adjacent basins. Each shoal is characterized by a length, width, and 
depth class, together with a Manning's friction coefficient that governs 
the resistance to flow. Initial shoal geometries were derived from 
nautical charts, aerial photography, and field observations of 
connectivity, recognizing that vegetative resistance varies over time and 
location. The initial 
estimates of shoal geometry were subsequently adjusted during calibration 
to bring modeled salinity into agreement with observations, as described in 
Section~\ref{sec:calibration}.

It is important to recognize that the BAM domain is a functional 
representation of Florida Bay rather than a geometric replica. Shoal 
dimensions and depth classes are not survey-derived measurements 
intended to precisely reproduce local bathymetry, they are calibrated 
parameters that encode the aggregate hydraulic behavior of the natural 
flow pathways they represent. The complexity of Florida Bay, including 
irregular bank topography, seasonal seagrass growth, and spatially variable 
sediment dynamics, cannot be resolved at the basin scale. Estimates are 
sufficient for the model's intended purpose and appropriate for the box
model design.

This design philosophy makes BAM particularly well suited for 
restoration scenario assessment. Because shoal connectivity and basin 
geometry have been calibrated against a 26-year observational record 
spanning a wide range of hydrologic conditions, the model encodes the 
aggregate sensitivity of Florida Bay salinity to changes in Everglades 
freshwater delivery and marine boundary forcing. Scenario projections 
that alter these boundary conditions, whether through upstream 
restoration actions or sea level rise, are therefore evaluated 
against a domain that has been demonstrated to reproduce the observed 
salinity response of the real system, within the limitations described 
in Section~\ref{sec:physical}.

\subsection{Physical Representation}
\label{sec:physical}

Following \citet{Park2016}, the physical basis of BAM is deliberately simple.
Mass transport across a shoal is computed independently within each 0.305 m
(1 ft) depth bin of the shoal cross section and summed over bins. Within a
bin the transport velocity $v$ is
\begin{equation}
   v = \sqrt{2 g \, \frac{h_u - h_d}{1 + f}}
   \label{equ:velocity}
\end{equation}
where $h_u$ and $h_d$ are the upstream and downstream heads referenced to that
bin's shoal crest and $g$ is gravitational acceleration, evaluated at the
domain latitude as $9.790\ \mathrm{m\,s^{-2}}$. Equation~\ref{equ:velocity} is
an energy formulation, not Manning's equation; shoal resistance enters through
the dimensionless friction factor
\begin{equation}
  f = 2 g \, n^{2} \, w \, R^{-4/3}
  \label{equ:friction}
\end{equation}
in which $n$ is the Manning roughness coefficient, $w$ the along-flow distance
across the shoal, and $R$ the hydraulic radius, approximated as the mean flow
depth on the assumption of a wide, shallow rectangular section.
Equation~\ref{equ:friction} follows from equating the Manning and
Darcy--Weisbach expressions for the same head loss, so $f$ carries Manning
roughness into an otherwise energy-based velocity. Because $f$ depends on $R$
while $R$ depends on the velocity head, the two are solved iteratively at each
timestep to a velocity tolerance of $10^{-4}\ \mathrm{m\,s^{-1}}$.

Flow across a shoal may also become critical. When the downstream head falls
below $h_c = 2 h_u / (3 + f)$ the discharge no longer depends on it, and $h_d$
is replaced by $h_c$ so that $v = \sqrt{2 g h_u / (3 + f)}$. The calibrated
roughness coefficients span $n = 0.05$ to $4.0$, one to two orders of magnitude
above values conventional for open channels, which reflects the role of $n$ as
the primary tuning parameter for a functional rather than geometric
representation of the shoal network. The module \texttt{hydro.py} contains all
hydraulic computations.

The mass flux of dissolved substances over each shoal $M$ ($\mathrm{g}$) is 
calculated as the product of the concentration of the substance in the water 
$C$ ($\mathrm{g \, m^{-3}}$) and the water mass flux $F$ ($\mathrm{m^{3} \, s^{-1}}$) 
across the shoal. The mass fluxes $M$ are summed around the boundary of each basin, 
the net mass flux over all shoals is multiplied by the time step and, along with 
other inputs and outputs of mass during the time step, added to the mass in the 
basin at the beginning of the time step. The new mass is divided by the water 
volume to estimate the concentration at the end of the time step.

Calibration is largely dependent on careful adjustment of Manning's coefficient 
to bring calculated salinity into agreement with observed data. Spatially 
explicit calibration refinements have been performed, on a limited basin-by-basin 
level, by adjusting the shoal depth bins to alter connectivity between basins. 
This is an iterative process, initially setting shoal depths based on physical 
observations of connectivity and vegetative resistance. Later, this connectivity 
is adjusted in the model so that the resultant salinity more closely aligns with 
observations. This shift in connectivity is an estimate, at best it adjusts 
results in the model to compensate for the model's course design.

A major caveat that influences calibration of the model is that the basins 
are considered well mixed at each time step. The calculated salinity represents 
a spatial and temporal average. In contrast, measured salinity used for 
calibration is a point measurement. It's reasonable to have differences in 
these two representations of salinity. Nonetheless, modeled results are 
strongly tied to observations and provide a reasonable means of projecting 
outcomes based on changes in stage and flow at the marsh boundary, even 
modeled stage from upstream restoration planning efforts.

\subsection{Boundary Forcing}

\begin{figure}[htbp]
  \centering
  \includegraphics[width=\textwidth]{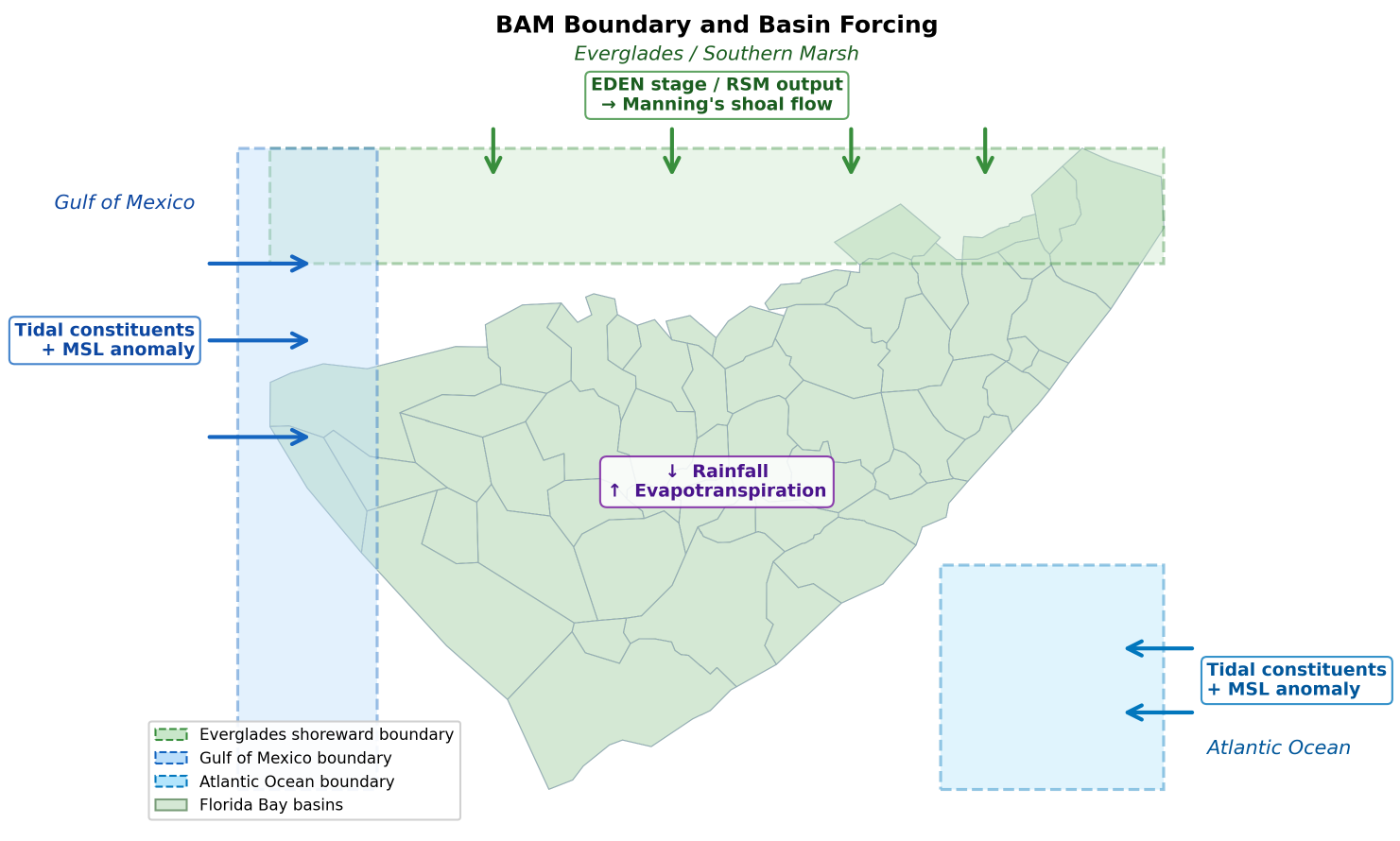}
  \caption{Schematic of BAM boundary and basin forcing. Arrows indicate the 
  direction of forcing inputs at each boundary. Along the shoreward (northern) 
  margin, water levels from the Everglades Depth Estimation Network (EDEN) or 
  upstream restoration planning models such as the Regional Simulation Model 
  (RSM) drive freshwater runoff into coastal basins via Manning's shoal 
  hydraulics. Along the marine margins, tidal boundary conditions derived from 
  NOAA subordinate station harmonic constituents, superimposed with a regional 
  mean sea level anomaly, are imposed at Gulf of Mexico and Atlantic Ocean 
  boundary basins. Each interior basin receives direct forcing from rainfall 
  and evapotranspiration.}
  \label{fig:forcing}
\end{figure}

BAM is driven by three categories of boundary forcing, each operating on 
different spatial and temporal scales (Figure~\ref{fig:forcing}). Along 
the northern boundary, where Florida Bay meets the southern reaches of 
Taylor Slough and the Everglades, the difference between water levels in the marsh 
and water levels in the neighboring bay are a critical driver. Water levels 
were extracted from the EDEN (\citealt{Telis2014}) at 
eight locations proximal to the coast. Freshwater runoff into the coastal 
basins is governed by the same shoal hydraulics used throughout 
the model. This formulation captures the combined contributions of streamflow, 
sheetflow, and groundwater exchange that characterizes Everglades discharge 
into Florida Bay, particularly along the northeastern margin \citep{Corbett1999, 
Hittle2001, Kelble2007}. Critically, EDEN stages can be replaced with output 
from upstream hydrologic models such as the Regional Simulation Model (RSM), 
making BAM a direct downstream linkage for assessing restoration scenarios.

Along the marine margin, tidal variability is the primary driver of exchange. 
Tidal water levels are computed from harmonic constituents at local NOAA 
subordinate stations, capturing all astronomical forcings including 
intraseasonal and interannual variations. A regional mean sea level anomaly 
was constructed from the NOAA tide gauges at Virginia Key, Vaca Key, and 
Key West, and superimposed on the tidal signal to account for secular sea 
level trends. Remaining ocean variability was not resolved by the harmonic 
constituents alone.

At the basin level, each of the 54 interior basins receives direct forcing 
from precipitation and evapotranspiration. Rainfall inputs are derived from 
the Everglades National Park marine monitoring network stations and aggregated 
to each basin using a weighted, spatially-smoothed, multi-gauge scheme. 
Evapotranspiration is applied as a single daily timeseries across the model
domain. Four exceptionally shallow basins, with amplified diurnal 
temperature ranges, additionally receive a Clausius--Clapeyron
amplification based on those observed daily maximum water temperature. The series is
derived from the GOES-based Everglades Depth Estimation Network potential
evapotranspiration product, whose record in the source database ends
2010-12-31 at every station. The period from 2011 onward is therefore
reconstructed rather than observed, and accounts for 57\% of the
evapotranspiration record. This large gap is filled using seasonal climatology of 
the observed period with a first-order autoregressive anomaly, which preserves 
the variance and short-term structure of the record but carries no information 
about the conditions of any particular day.

Boundary salinities on the Gulf and Atlantic margins are specified from 
observed monitoring data and exert an important influence on interior basin salinities, 
particularly during periods of weak freshwater forcing.

\section{Observational Data and Model Inputs}
\label{sec:data}

\subsection{Marine Monitoring Network}
\label{sec:mmn}

The Everglades National Park marine monitoring network is the primary 
source of hydrographic data across Florida Bay. Established over several 
years, beginning in 1989, the network contains 17 monitoring stations, 
each capable of measuring multiple parameters at hourly or sub-hourly time 
steps and delivering that data via NOAA's GOES satellite system. This live 
data stream includes hourly precipitation and salinity, the latter derived 
from temperature and conductivity measured at a nominal 30.5 cm (1 ft.) 
above the bay bottom. Each station also measures water level on 6 min. time 
steps to an accuracy of 0.006 m (0.02 ft) relative to NAVD88. Maintenance 
and calibration are performed monthly, on site, through a process of checking 
values against a laboratory calibrated standard for salinity and through 
manual measurements for water level. Instruments are cleaned, calibrated, and 
redeployed on each site visit and data is corrected for any instrument drift 
through post-processing within the database.

Gaps in the observational record are filled by seasonal quantile mapping from
the most strongly correlated companion station, a nonparametric distribution
transformation of the kind reviewed by \citet{Gudmundsson2012}. For each missing
day the donor value is converted to its quantile within the donor's empirical
distribution for that season, and the target value at the same quantile is
returned, so the fill inherits the donor's day-to-day sequencing while
reproducing the target station's own distribution. The filled segment is offset
to agree with the observations immediately either side of the gap, the offset
decaying into the interior at the station's own autocorrelation timescale, and
its day-to-day variability is matched to that of the station where it would
otherwise be too rough. Where no companion station is available,
seasonal climatology is used with a first-order autoregressive anomaly bridged
between the gap edges, which preserves the variance and short-term structure of
the record.

Gaps longer than one year are not filled, and as such, 
the Murray Key water level record, lost to Hurricane
Irma in 2017, is left absent rather than reconstructed. Model performance in
Section~\ref{sec:performance} is evaluated against observed days only, so no
resampled value enters any reported skill metric. Station locations and the
variables collected at each are shown in Figure~\ref{fig:domain}, and rainfall
timeseries for eastern and western Florida Bay stations are illustrated in the
model manual \citep{Park2016}. The complete set of model input files, their
periods of record, and the command-line options that select them are listed in
Table~\ref{tab:inputs}.

\begin{table}[htbp]
\caption{BAM input data files, period of record, and command-line options. 
Salinity and Stage inputs are used for model output comparison and 
initialization, not as explicit model forcings, though salinity timeseries 
can be imposed on basins with the \texttt{-{}-gaugeSalinity} option. 
$^a$The Sea Level record, released monthly, extends to 2026-05-15
to anchor the interpolating spline past the end of the model run.}
  \label{tab:inputs}
  \centering
  \begin{tabular}{llllr}
    \toprule
    Variable  & File & Start & End & Option \\
    \midrule
    Rain      & \texttt{DailyRainFilled\_cm\_1999-9-1\_2026-4-30.csv}
              & 1999-09-01 & 2026-04-30 & \texttt{-br} \\
    ET        & \texttt{PET\_1999-9-1\_2026-4-30.csv}
              & 1999-09-01 & 2026-04-30 & \texttt{-et} \\
    Tide      & \texttt{Basin\_Tide\_Boundary\_2000\_2026.csv}
              & 1999-09-01 & 2026-04-30 & \texttt{-bt} \\
    Sea Level & \texttt{MSL\_Anomaly.csv}
              & 1999-08-15 & 2026-05-15$^a$ & \texttt{-msl} \\
    Runoff    & \texttt{EDEN\_Stage\_OffsetMSL\_1999-9-1\_2026-4-30.csv}
              & 1999-09-01 & 2026-04-30 & \texttt{-bR} \\
    Flow      & \texttt{S197\_Flow\_1999-9-1\_2026-4-30.csv}
              & 1999-09-01 & 2026-04-30 & \texttt{-bc} \\
    Salinity  & \texttt{DailySalinityFilled\_1999-9-1\_2026-4-30.csv}
              & 1999-09-01 & 2026-04-30 & \texttt{-sf} \\
    Stage     & \texttt{DailyStage\_1999-9-1\_2026-4-30.csv}
              & 1999-09-01 & 2026-04-30 & \texttt{-bs} \\
    \bottomrule
  \end{tabular}
\end{table}

\subsection{Tidal and Sea Level Forcing}
\label{sec:tides}

Tidal variations are a primary driver of hydraulic fluxes throughout Florida
Bay, particularly along the Gulf of Mexico and Atlantic Ocean margins. BAM
computes tidal boundary conditions from harmonic constituents at seven local
NOAA subordinate stations distributed around the bay
(Table~\ref{tab:tide_stations}), following standard tidal analysis and
prediction practice \citep{Parker2007} and capturing all astronomical forcings
including intraseasonal and interannual variations. Hourly tidal timeseries
generated from these constituents are interpolated to the model timestep at
runtime using cubic spline fits. The tidal boundary data used in this effort
spans 1999 to 2026, to encompass the model period of record.

The Gulf boundary of the BAM domain proved particularly challenging given 
that there are no tidal stations along this transect. Instead, there are two 
stations, one on Cape Sable and the other at Long Key, showing substantially 
different tides. Based on the harmonic analysis of available station records,
Long Key is delayed behind Cape Sable by 6 to 8 hours and the 
magnitude of tidal cycles is attenuated by a factor of 2.7 to 3.0. Tidal 
forcing for the Gulf boundary is then estimated by amplitude and phase interpolation 
between the two stations, with the Gulf region assigned 2/3 the Cape Sable
amplitude with a 3-hour delay, and the southern Gulf region assigned 4/3 of 
the Long Key amplitude applied with a 3-hour lead. 

\begin{figure}[htbp]
  \centering
  \includegraphics[width=\textwidth]{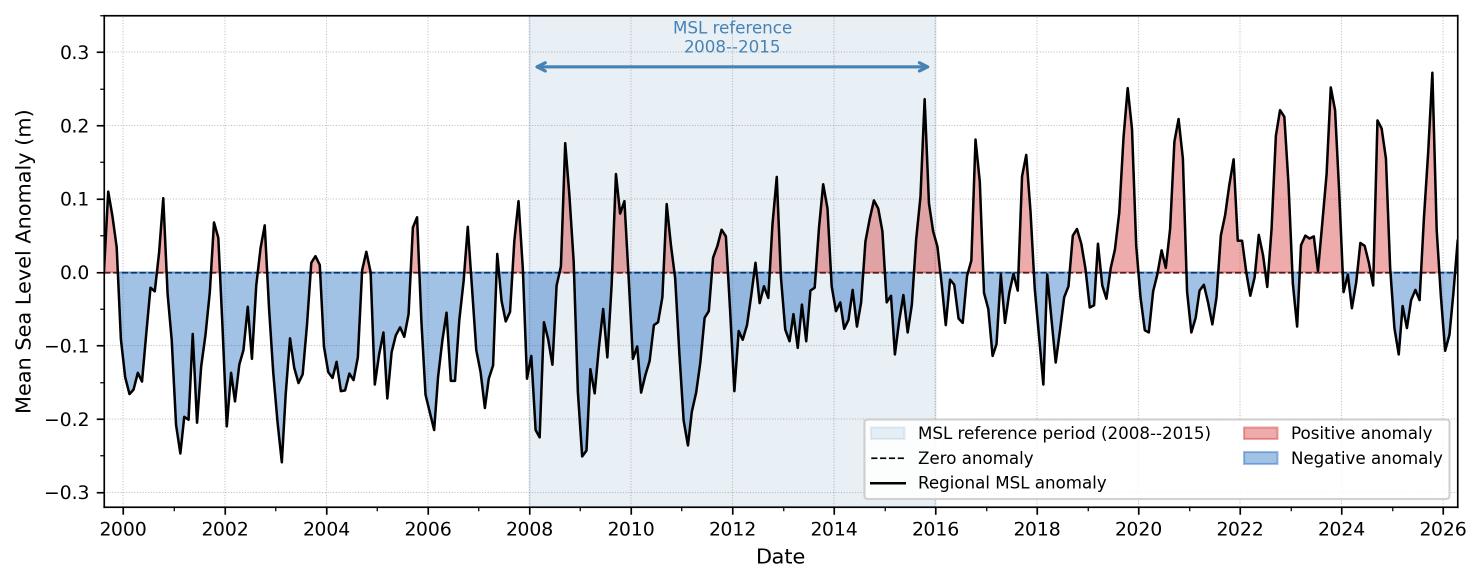}
\caption{Regional mean sea level anomaly with respect to the 2008--2015 
mean sea level in Florida Bay ($-$14.8 cm NAVD88). The anomaly is 
constructed from monthly mean sea level data at three long-term primary 
NOAA tide gauges (Virginia Key, Vaca Key, and Key West) and spans the 
full model period of record 1999--2026. The shaded band indicates the 
reference period used to define mean sea level in Florida Bay. Positive 
anomalies (red) indicate periods of elevated regional sea level; negative 
anomalies (blue) indicate suppressed sea level.}
  \label{fig:msl}
\end{figure}

Additionally, a regional sea level anomaly timeseries was superimposed on the 
tidal boundary conditions for all the Gulf and southern boundary basins. This 
anomaly is constructed from monthly 
mean sea level data at three long-term NOAA tide gauges: Virginia Key, Vaca 
Key, and Key West (Table~\ref{tab:noaa_gauges}). A single regional timeseries is 
created by adding the annual seasonal cycle to the published monthly mean sea levels,
which are converted to the NAVD88 
datum and averaged across the three stations
(root mean square deviation of 1.7 cm among stations). The resulting timeseries is then 
referenced to the mean sea level in Florida Bay over the 2008--2015 period, 
consistent with the anomaly reference frame used throughout the model. This regional 
mean sea level anomaly timeseries extends from 1999-08-15 to 2026-05-15 
(Figure~\ref{fig:msl}). The Florida Bay mean sea level reference of $-$14.8 cm NAVD88 is tied 
to the NOAA National Tidal Datum Epoch (NTDE) of 1983--2001 and will 
require periodic updating as appropriate.

\begin{table}[htbp]
  \caption{NOAA subordinate tide stations used to compute tidal boundary 
  conditions for BAM Gulf of Mexico and Atlantic Ocean boundary basins.}
  \label{tab:tide_stations}
  \centering
  \begin{tabular}{ll}
    \toprule
    Station Name & Station ID \\
    \midrule
    Cape Sable, East Cape, Florida          & TEC4165   \\
    Long Key, western end, Florida          & 8723899   \\
    Lignumvitae Key, NE side, Florida Bay   & 8723824   \\
    Snake Creek, Hwy.\ 1 bridge, Windley Key & 8723787  \\
    Tavernier Creek, Hwy.\ 1 bridge, Hawk Channel & 8723748 \\
    Garden Cove, Key Largo, Florida         & 8723622   \\
    Little Card Sound bridge, Florida       & 8723534   \\
    \bottomrule
  \end{tabular}
\end{table}

\begin{table}[htbp]
  \caption{Long-term primary NOAA tide gauges used to construct the regional 
  mean sea level anomaly timeseries. MSL is the NTDE mean sea level datum 
  elevation and NAVD the North American Vertical Datum elevation,
  as reported in \citep{Park2016}, updated to the extended period of record.}
  \label{tab:noaa_gauges}
  \centering
  \begin{tabular}{lllrr}
    \toprule
    Station & CO-OPS ID & Verified Data & NAVD (m) & MSL (m) \\
    \midrule
    Virginia Key & 8723214 & 1994--2026 & 3.698 & 3.431 \\
    Vaca Key     & 8723970 & 1973--2026 & 1.182 & 0.931 \\
    Key West     & 8724580 & 1954--2026 & 1.928 & 1.662 \\
    \bottomrule
  \end{tabular}
\end{table}

\subsection{Everglades Stage and Runoff}

Water levels in the southern Everglades and freshwater runoff into the coastal 
basins of Florida Bay have a strong negative correlation that drive coastal 
basin salinity \citep{Tabb1967, Kelble2007}. BAM uses the existing head gradient 
to determine runoff, including streamflow, sheetflow and groundwater exchange, 
into the coastal basins. This approach 
avoids the need for explicit flow measurements at the marsh-bay interface, which 
are sparse and difficult to maintain, and instead relies on the more readily 
available and spatially continuous water level data from the EDEN (\citealt{Telis2014}).

For observation based calculations, EDEN provides interpolated daily water level
surfaces in the southern Everglades region at 400 m resolution. During calibration, 
EDEN stage values were extracted along the shoreward boundary from the western gulf 
coast to northeastern Florida Bay. Water levels were converted to NAVD88 as 
described above and then used to drive runoff into the corresponding coastal 
basin through the calibrated shoals. Shoal properties having been adjusted to 
match aggregate runoff from the FATHOM model \citep{Cosby2010} over the overlapping 
period of record.

A key strength of this approach is that EDEN stages can be replaced directly with 
output from upstream hydrologic models. Restoration planning efforts in the Everglades 
routinely produce water level projections using tools such as the RSM or the 
South Florida Water Management Model (SFWMM). By substituting 
RSM or SFWMM stage output for EDEN observations at the eight shoreward boundary 
stations, BAM provides a direct and computationally efficient linkage between upstream 
restoration scenarios and their salinity consequences in Florida Bay. This capability 
is central to the intended application of BAM as a scenario assessment tool in the 
context of the Comprehensive Everglades Restoration Program (CERP).

\begin{figure}[htbp]
  \centering
  \includegraphics[width=\textwidth]{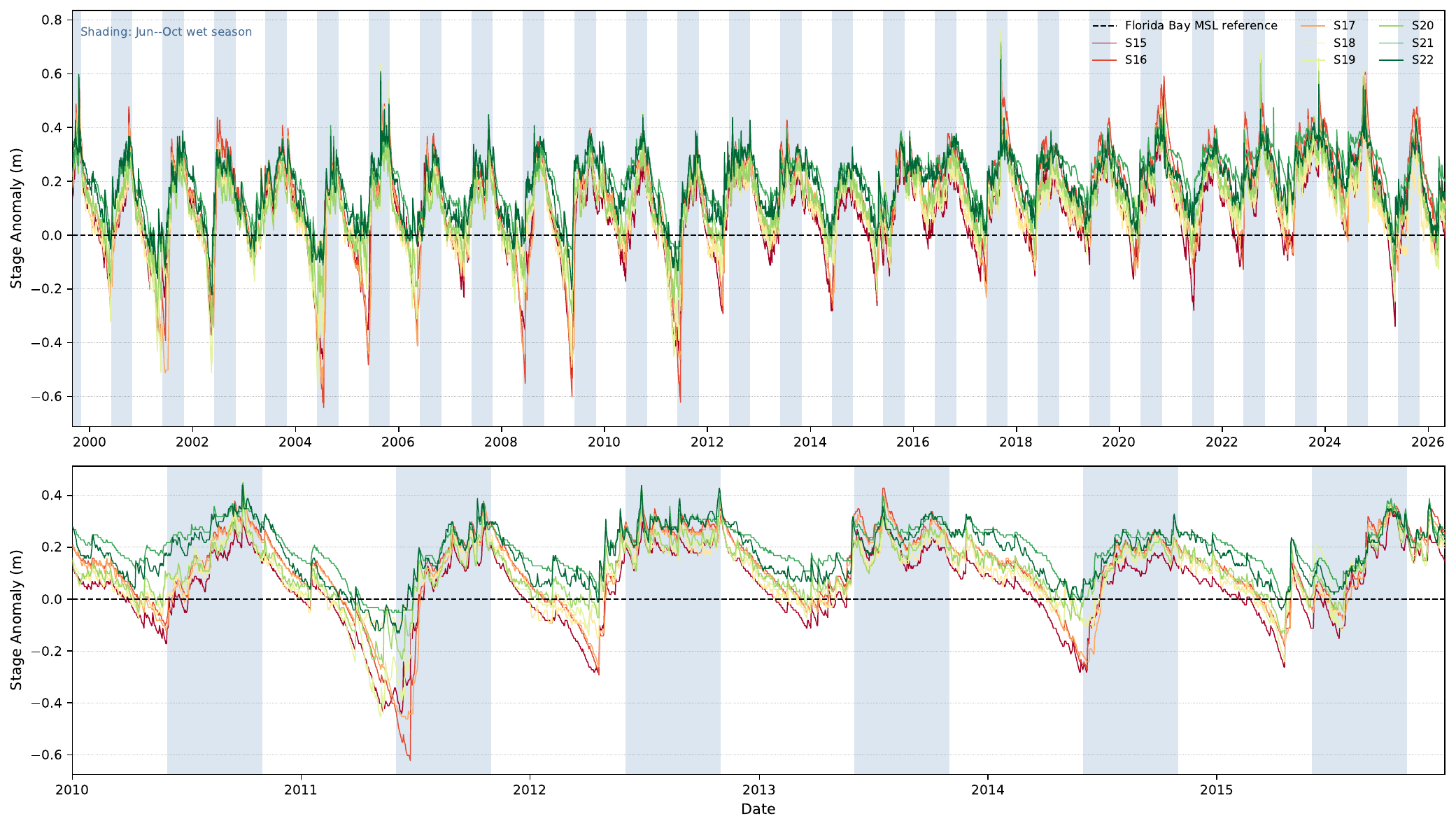}
  \caption{Daily EDEN stage anomalies at eight stations in the southern 
  Everglades relative to the Florida Bay mean sea level reference 
  ($-$14.8 cm NAVD88), spanning the full model period of record 
  1999--2026. Station colors progress from west (S15, red-yellow) to 
  east (S22, green), illustrating the spatial gradient in stage across 
  the shoreward boundary. Blue shading marks the June--October wet 
  season. The lower panel shows the 2010--2015 period in detail, 
  highlighting the seasonal cycle and interannual variability that 
  drive freshwater runoff into Florida Bay coastal basins.}
  \label{fig:eden}
\end{figure}

The EDEN stage file used in this effort spans the full model period of record, 
1999-09-01 to 2026-04-30. Water levels at the eight boundary stations
(Table~\ref{tab:eden_stations}) exhibit the expected strong seasonal signal,
with stages rising through the wet season
(June--October) and falling to annual minima during the dry season
(November--May), as well as substantial interannual variability associated with 
drought and high-water years (Figure~\ref{fig:eden}).

\begin{table}[htbp]
  \caption{EDEN stations providing Everglades stage to drive freshwater 
  runoff into Florida Bay coastal basins. Coordinates are UTM Zone 17N 
  (NAD83) as reported in \citep{Park2016}, updated to the extended period of record.}
  \label{tab:eden_stations}
  \centering
  \begin{tabular}{lrrrr}
    \toprule
    Station & UTM North (m) & UTM East (m) & Longitude & Latitude \\
    \midrule
    S15 & 2{,}794{,}000 & 520{,}000 & $-$80.801 & 25.262 \\
    S16 & 2{,}792{,}500 & 525{,}000 & $-$80.752 & 25.249 \\
    S17 & 2{,}790{,}000 & 530{,}000 & $-$80.702 & 25.226 \\
    S18 & 2{,}791{,}500 & 535{,}000 & $-$80.652 & 25.239 \\
    S19 & 2{,}794{,}500 & 540{,}000 & $-$80.603 & 25.266 \\
    S20 & 2{,}795{,}000 & 545{,}000 & $-$80.553 & 25.271 \\
    S21 & 2{,}795{,}500 & 550{,}000 & $-$80.503 & 25.275 \\
    S22 & 2{,}795{,}700 & 555{,}200 & $-$80.452 & 25.277 \\
    \bottomrule
  \end{tabular}
\end{table}

\FloatBarrier

\section{Calibration}
\label{sec:calibration}

Calibration of BAM is described in detail in the model manual \citep{Park2016}. 
Briefly, the primary calibration parameter is the Manning's coefficient applied 
to each shoal, adjusted iteratively to bring modeled salinity into alignment with 
observations from the Marine Monitoring Network. Secondary calibration was performed 
on a limited basin-by-basin basis by adjusting shoal depth bins to modify 
interbasin connectivity. Shoal properties along the Everglades shoreward boundary 
were calibrated to match aggregate runoff from the FATHOM model \citep{Cosby2010} 
over the overlapping rainy seasons of 2001--2003. No changes to calibration 
parameters were made during the extension of the period of record to 2026; 
rather, the multi-decade performance evaluation presented in 
Section~\ref{sec:performance} serves as a post-hoc validation of this 
original calibration, testing whether parameters tuned against a three-year 
overlap with FATHOM remain adequate across the full diversity of hydrologic 
conditions encountered over the subsequent two decades.

\section{Model Performance}
\label{sec:performance}

Evaluation of BAM performance over the full period of record focuses on two 
primary state variables: water level and salinity. Water level simulation is addressed 
first, as accurate interbasin hydraulic gradients are a prerequisite for correct 
mass transport and therefore directly govern the quality of salinity projections. 
Salinity performance, which is the ultimate measure of model utility for restoration 
planning, is addressed in Section~\ref{sec:salinity_performance}. Model bias
and an evaluation of the model's performance over the extended period
of record run are also included in this evaluation.

\subsection{Water Level Simulation}
\label{sec:stage_performance}

Accurate simulation of water levels across Florida Bay is fundamental to BAM's
utility as a restoration planning tool. The hydraulic gradients that drive
interbasin flow, and therefore salinity transport, are determined entirely by
water level differences across shoals. A model that captures the timing and
magnitude of water level variability across the bay will correctly partition
freshwater and marine influences on individual basins, providing credible
salinity projections under altered upstream conditions.

Water levels in BAM are referenced as anomalies with respect to a zero shoal
depth datum rather than an absolute geodetic reference. When substituting water
levels from an upstream hydrologic model, such as RSM, for EDEN observations at
the shoreward boundary, the external stage timeseries must be converted to the
BAM anomaly reference frame before use. The conversion is done by subtracting 
the Florida Bay mean sea level value ($-$14.8 cm NAVD88,
Section~\ref{sec:tides}) from the external stage timeseries. If the upstream
model reports water levels in NAVD88 meters, the offset is applied as:

\begin{equation}
  h_{BAM} = h_{NAVD88} - h_{ref}
  \label{equ:stage_offset}
\end{equation}

\noindent where $h_{BAM}$ is the anomaly-referenced stage supplied to BAM (m), 
$h_{NAVD88}$ is the upstream model water level in NAVD88 (m), and $h_{ref}$ is 
the Florida Bay mean sea level reference ($-$0.148 m NAVD88). If the upstream 
model reports water levels in a different datum, such as NGVD29, an additional 
datum conversion must be applied prior to the offset subtraction using the 
appropriate regional datum transformation. In South Florida, the NGVD29 to 
NAVD88 conversion varies spatially from approximately $-$0.15 m to $-$0.46 m 
and should be applied using NOAA's VDatum tool or equivalent regional 
transformation rather than a single constant. BAM ingests stage boundary 
conditions as daily values in meters, provided as comma-delimited ASCII files 
consistent with the format described in the model manual \citep{Park2016}.

\begin{figure}[htbp]
  \centering
  \includegraphics[width=\textwidth]{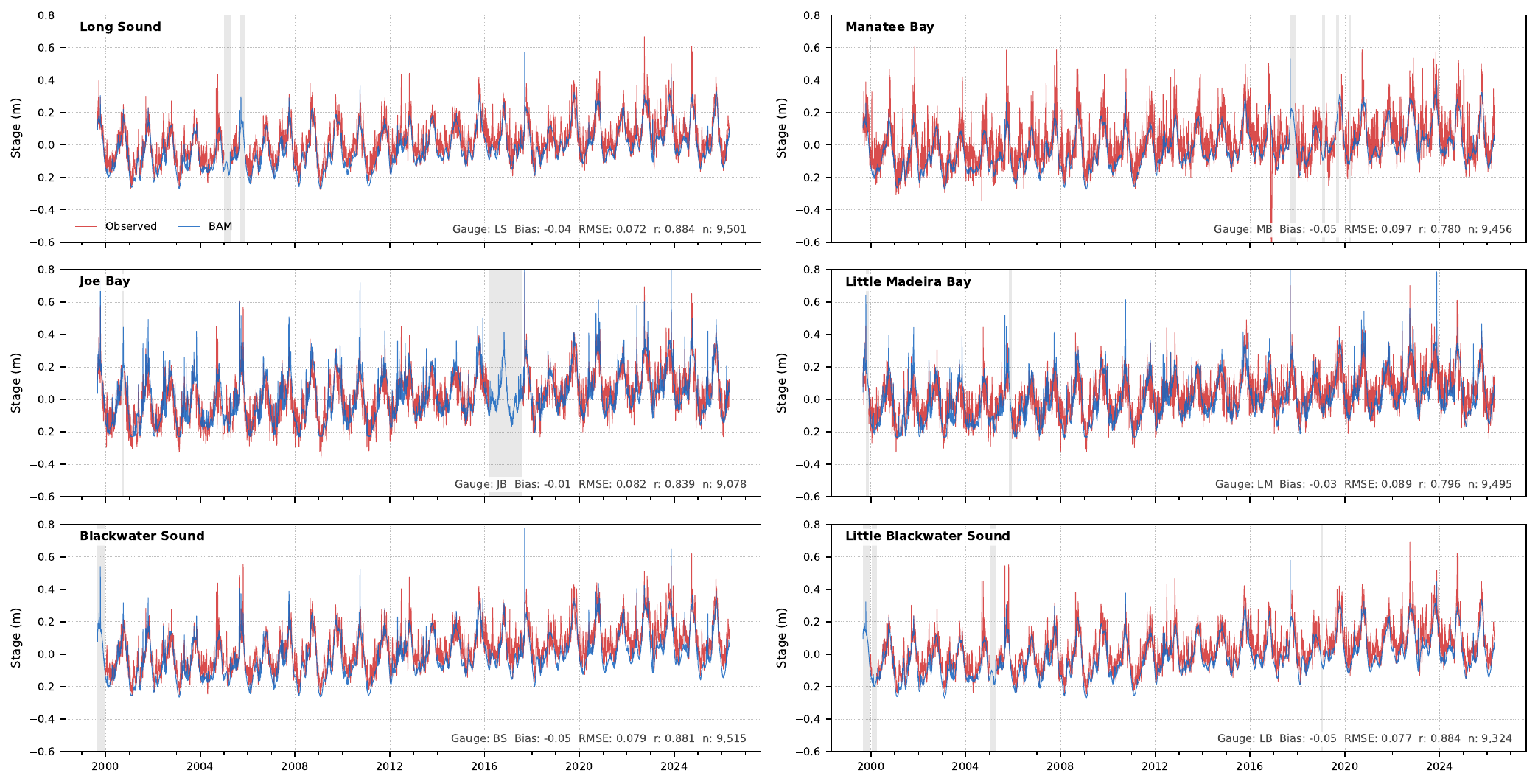}
  \caption{Comparison of BAM-computed water levels (blue) and observed Marine
  Monitoring Network stage data (red) for northeastern Florida Bay basins,
  1999--2026. Observations are plotted for observed days only; grey bands mark
  periods of 30 or more consecutive days without observations. Basins in this
  region receive direct freshwater forcing from the Everglades shoreward
  boundary and show strong agreement between modeled and observed water level
  variability.}
  \label{fig:stage_ne}
\end{figure}

\begin{figure}[htbp]
  \centering
  \includegraphics[width=\textwidth]{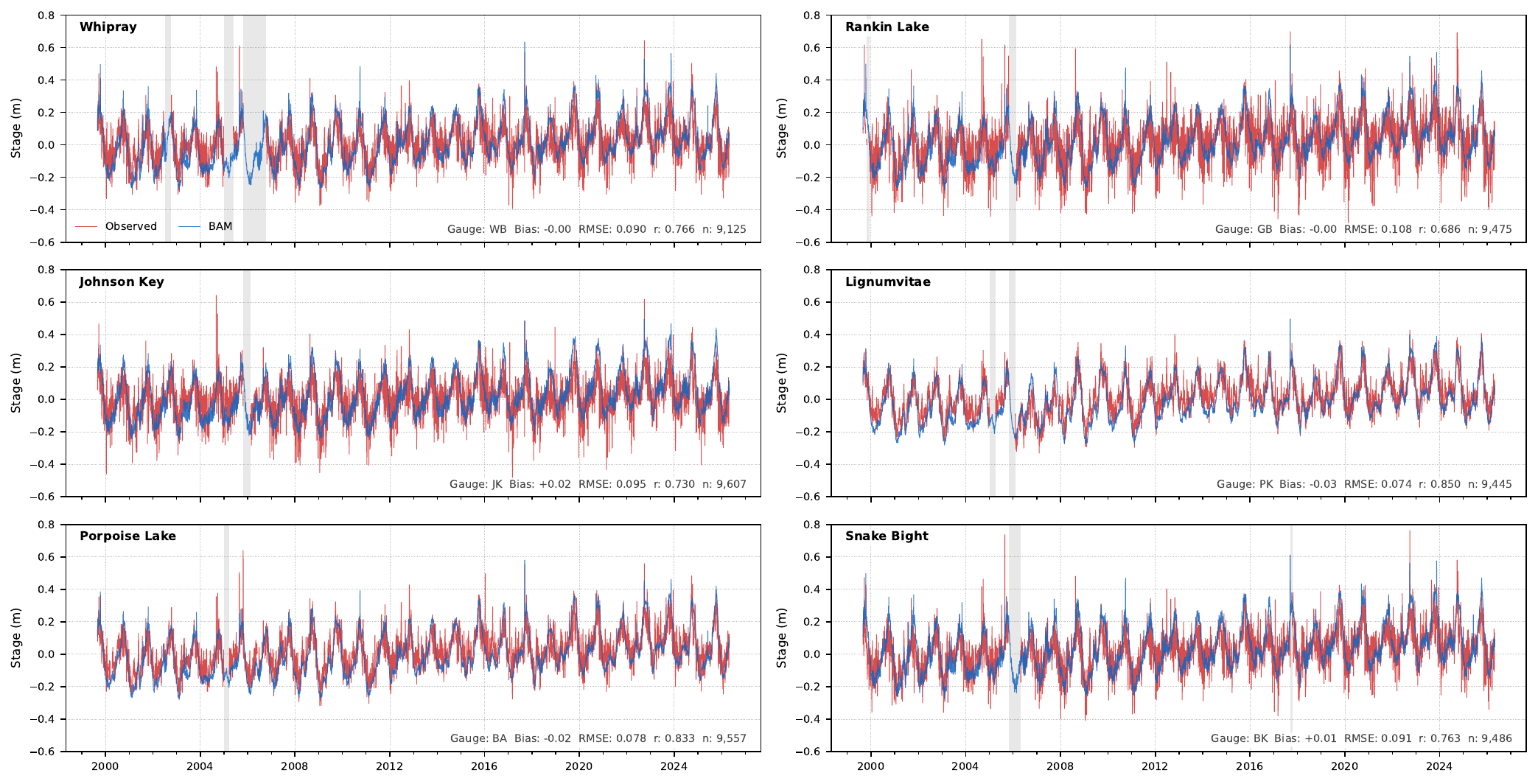}
  \caption{Comparison of BAM-computed water levels (blue) and observed Marine
  Monitoring Network stage data (red) for central and western Florida Bay
  basins, 1999--2026. Grey bands mark periods of 30 or more consecutive days
  without observations. These basins are more isolated from direct Everglades
  forcing and more sensitive to shoal connectivity assumptions, resulting in
  greater variability in model-observation agreement relative to the
  northeastern bay.}
  \label{fig:stage_cw}
\end{figure}

The comparison of modeled and observed water levels presented here is qualitative 
in terms of absolute magnitude but meaningful in terms of temporal variability, 
phase, and the relative response of basins across the domain. 
Figures~\ref{fig:stage_ne} and \ref{fig:stage_cw} compare BAM-computed water 
levels against Marine Monitoring Network observations for northeastern and 
central-western Florida Bay basins over the full model period of record 
1999--2026. The model captures the dominant tidal and seasonal signals across 
the domain, with basin water levels responding correctly to both marine tidal 
forcing along the Gulf and Atlantic boundaries and to freshwater stage 
fluctuations along the Everglades margin.

Over the full 1999--2026 period, domain-wide mean bias is $-$0.024 m, mean RMSE
is 0.092 m, and mean Pearson correlation is $r = 0.797$ across the 15 basins
with observational records. Basins in the northeastern bay show the strongest
agreement, with Little Blackwater Sound and Long Sound achieving the highest
correlations ($r = 0.884$ for both) and Long Sound recording an RMSE of 0.072 m
against a bias of $-$0.044 m. Basins in the central and western bay show greater
variability in model-observation agreement, consistent with their greater
hydrologic isolation and increased sensitivity to shoal connectivity
assumptions. Whipray, a representative central bay station, achieves
$r = 0.766$ with an RMSE of 0.090 m and a near-zero bias of $-$0.002 m. Rankin
Lake shows the lowest correlation ($r = 0.686$) among all stations. Deer Key
carries the largest bias in the domain at $-$0.117 m, reflecting the offset
between that basin and the TC gauge used to represent it. The near-zero
domain-wide mean bias 
confirms that the datum offset conversion described in 
Equation~\ref{equ:stage_offset} places modeled and observed water levels on 
a consistent reference frame across the full period of record, supporting the 
use of BAM output as a direct input to salinity projection under altered 
upstream boundary conditions.

\begin{figure}[htbp]
  \centering
  \includegraphics[width=\textwidth]{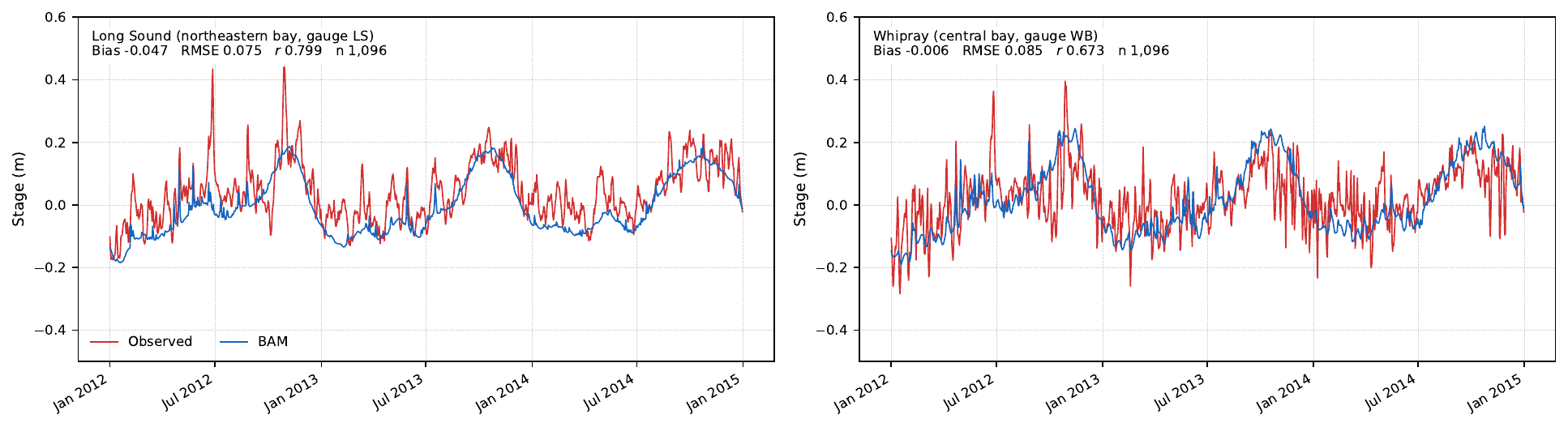}
  \caption{Detail water level comparison at Long Sound (northeastern bay, gauge
  LS) and Whipray (central bay, gauge WB) for the period 2012--2014. Skill
  metrics shown in each panel (bias, RMSE, and Pearson correlation coefficient
  $r$) are computed over the 2012--2014 detail period only and against observed
  days; full period of record metrics are reported in the text.}
  \label{fig:stage_detail}
\end{figure}

Figure~\ref{fig:stage_detail} shows a three-year detail comparison at Long 
Sound and Whipray for visual inspection of model-observation agreement at 
daily resolution.

\FloatBarrier

\subsection{Salinity Simulation}
\label{sec:salinity_performance}
\begin{figure}[htbp]
  \centering
  \includegraphics[width=\textwidth]{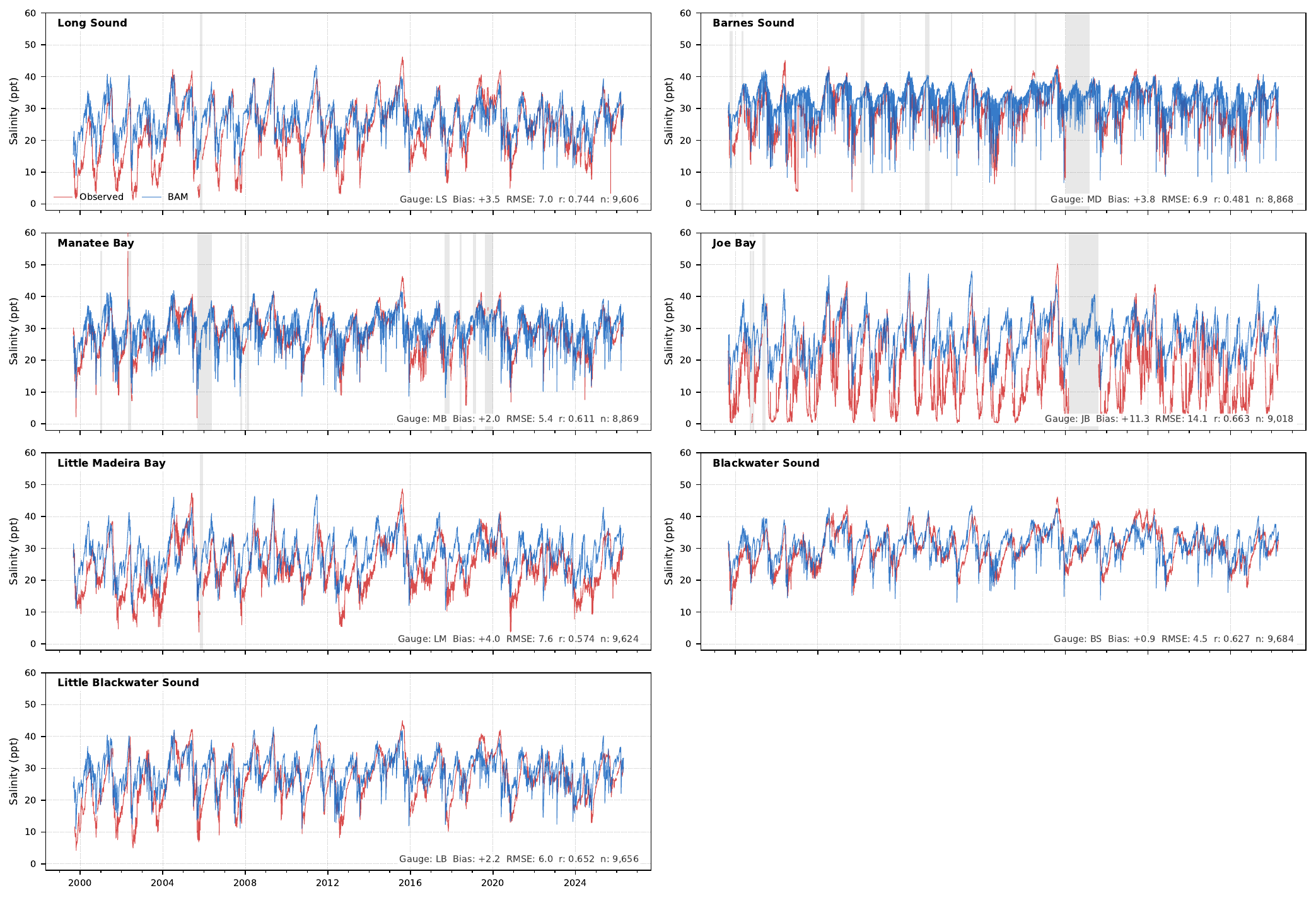}
  \caption{Salinity comparisons for northeastern Florida Bay basins, 1999--2026.
  Observed Marine Monitoring Network data (red) and BAM output (blue).
  Observations are plotted for observed days only; grey bands mark periods of 30
  or more consecutive days without observations, and skill metrics are computed
  over observed days across the full period of record. Northeast bay basins show
  consistent positive biases reflecting the well-mixed basin assumption and
  possible underestimation of freshwater delivery at the shoreward boundary. Joe
  Bay ($+$11.3 ppt bias) is discussed in the text.}
  \label{fig:salinity_ne}
\end{figure}

\begin{figure}[htbp]
  \centering
  \includegraphics[width=\textwidth]{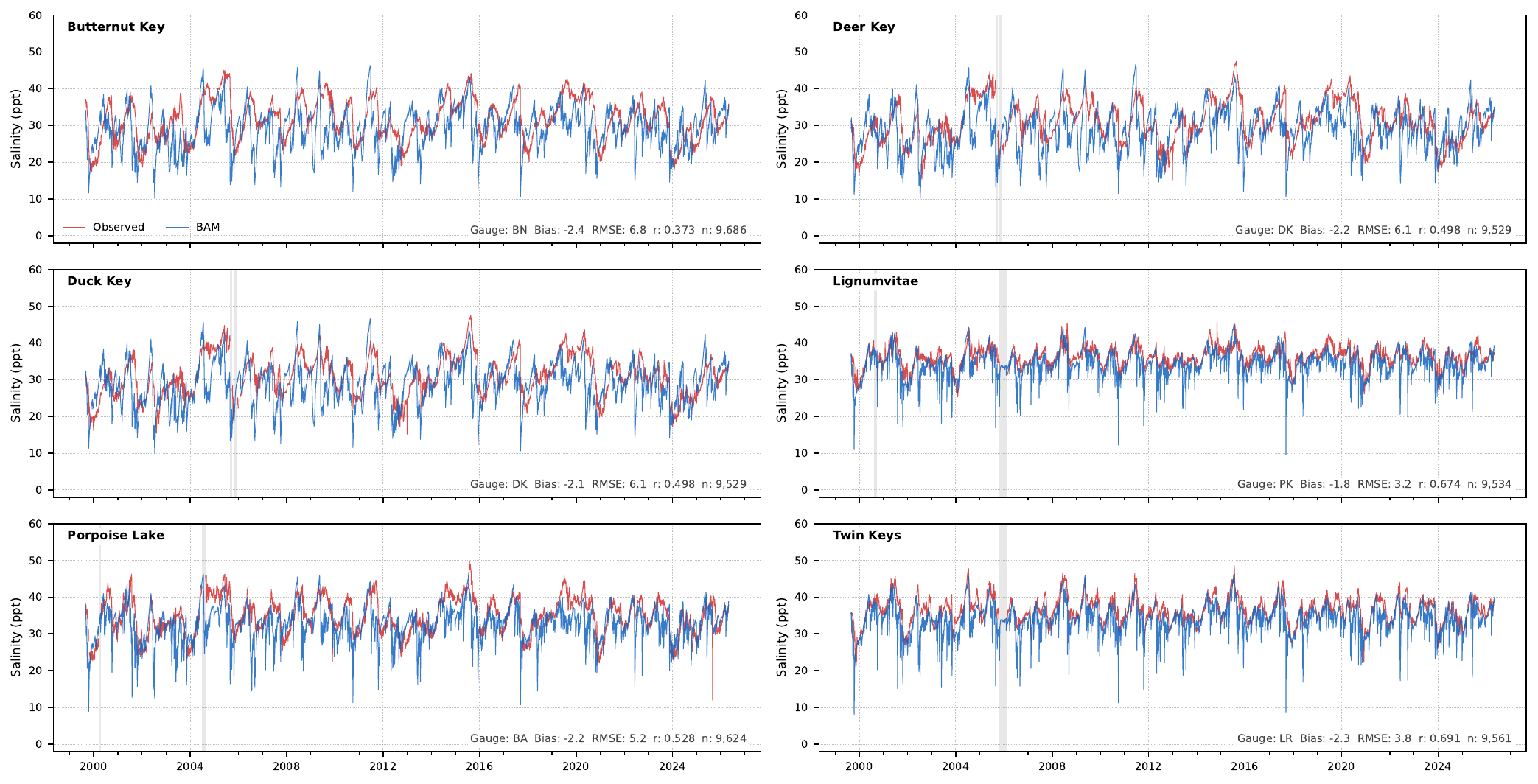}
  \caption{Salinity comparisons for central Florida Bay basins, 1999--2026.
  Observed Marine Monitoring Network data (red) and BAM output (blue). Grey
  bands mark periods of 30 or more consecutive days without observations.
  Central bay basins show small negative biases consistent with slightly excess
  marine connectivity in the model shoal network. Deer Key and Duck Key share
  the DK observational gauge; Twin Keys and Rabbit Key share the LR gauge.}
  \label{fig:salinity_central}
\end{figure}

\begin{figure}[htbp]
  \centering
  \includegraphics[width=\textwidth]{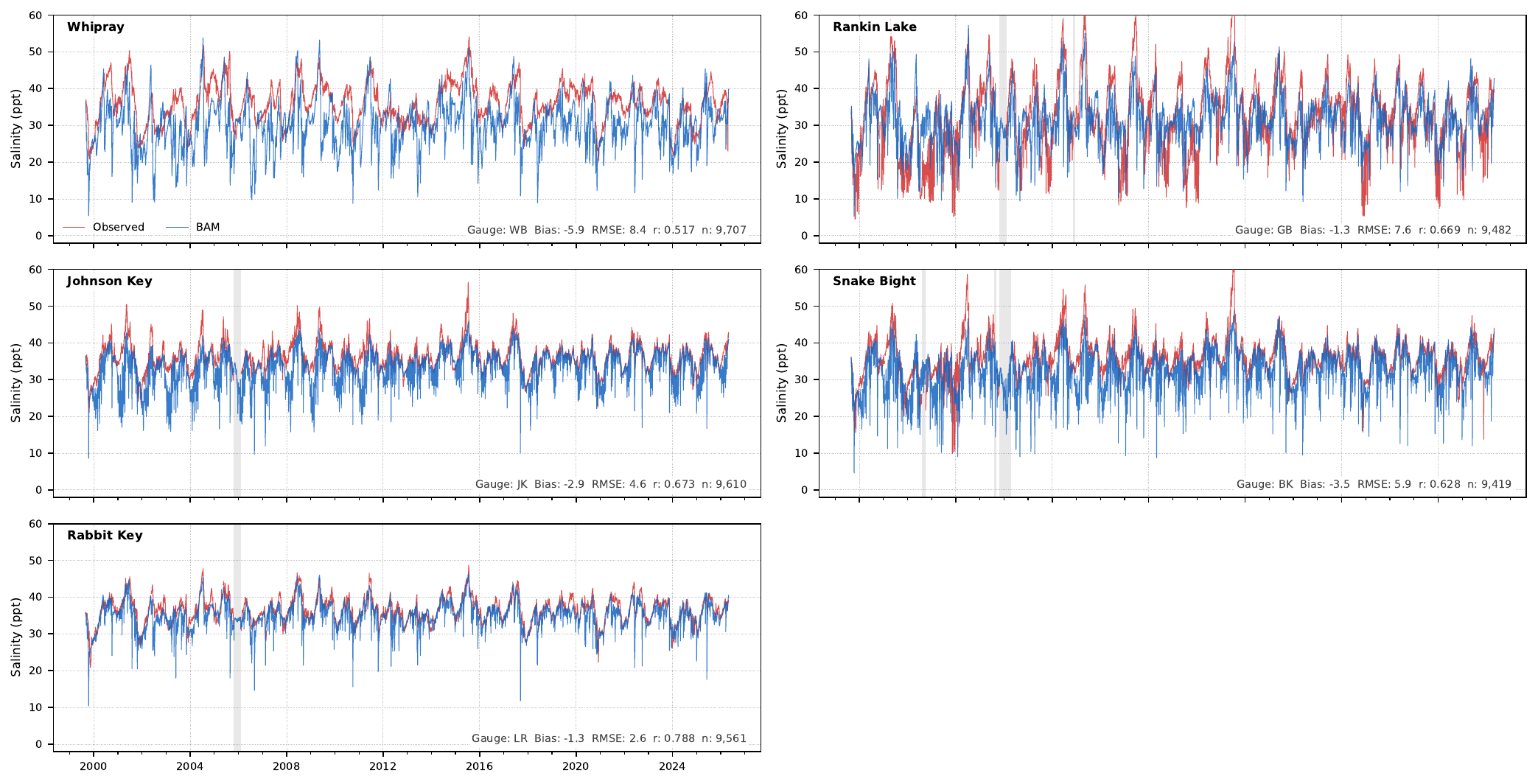}
  \caption{Salinity comparisons for western Florida Bay basins, 1999--2026.
  Observed Marine Monitoring Network data (red) and BAM output (blue). Grey
  bands mark periods of 30 or more consecutive days without observations.
  Whipray shows the largest negative bias ($-$5.9 ppt) in the domain. Rabbit Key
  and Rankin Lake show the strongest correlations in the western group
  ($r =   0.788$ and $r = 0.669$ respectively).}
  \label{fig:salinity_west}
\end{figure}

\begin{figure}[htbp]
  \centering
  \includegraphics[width=\textwidth]{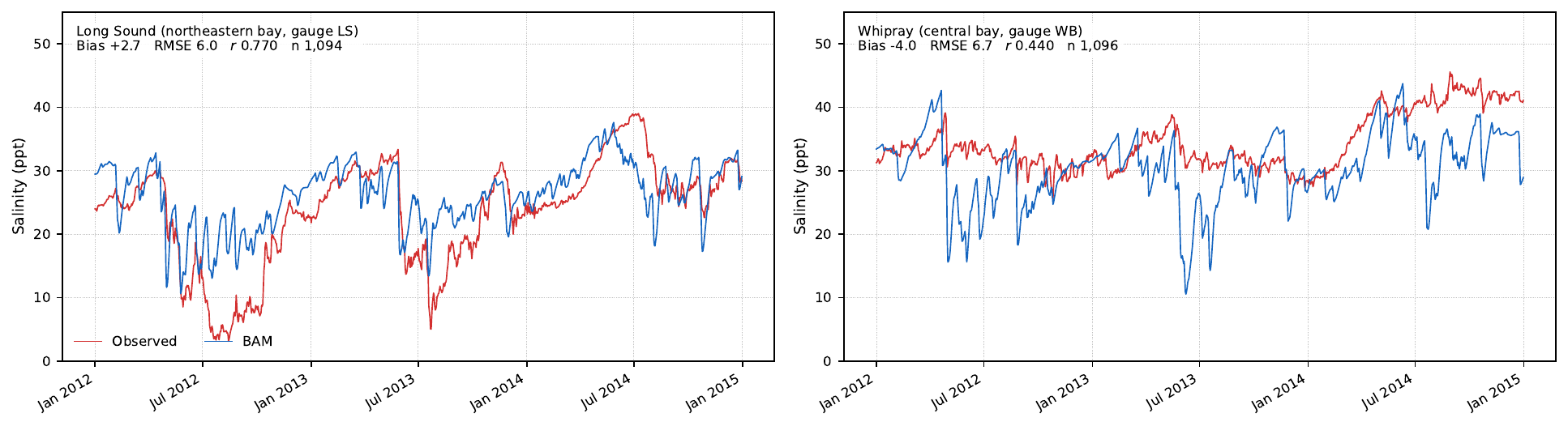}
  \caption{Detail salinity comparison at Long Sound (northeastern bay, gauge LS)
  and Whipray (central/western bay, gauge WB) for the period 2012--2014. Skill
  metrics shown in each panel are computed over the 2012--2014 detail period
  only and against observed days; full period of record metrics are reported in
  Table~\ref{tab:sal_skill}. Long Sound illustrates the strong seasonal
  freshwater signal characteristic of the northeastern bay; Whipray illustrates
  the more marine-dominated, lower-amplitude salinity regime of the central
  bay.}
  \label{fig:salinity_detail}
\end{figure}

Salinity prediction is the primary use of BAM, it is the principal metric 
by which model utility for restoration planning is assessed. Salinity 
comparisons for all 18 basins with observational records are presented 
in Figures~\ref{fig:salinity_ne}, \ref{fig:salinity_central}, and 
\ref{fig:salinity_west}, with quantitative skill metrics summarized in 
Table~\ref{tab:sal_skill}. A detail comparison at Long Sound and Whipray 
over 2012 to 2014, selected as a sample of the full hydrologic regime, 
is shown in Figure~\ref{fig:salinity_detail}.

Over the full 1999--2026 period of record the domain-wide mean bias is
0.00 ppt, mean RMSE is 6.21 ppt, and mean Pearson correlation is
$r = 0.605$ across 18 basins. The near-zero domain mean reflects cancellation
between opposing regional biases rather than an absence of bias: individual
basin biases span $-$5.9 to $+$11.3 ppt, seven positive and eleven negative,
giving a mean absolute bias of 3.1 ppt. These aggregate metrics reflect a
physically coherent spatial pattern in model performance. The Northeastern bay
basins show consistent positive biases 
ranging from $+$0.9 to $+$4.0 ppt (excluding Joe Bay), indicating that 
BAM delivers slightly too little freshwater to the shoreward margin in 
that region. Central and western bay basins show consistent negative 
biases of $-$1.3 to $-$5.9 ppt, reflecting a tendency toward 
lower than observed salinity in the more marine-influenced and also more 
isolated section of the bay. This opposing 
bias structure suggests a systematic partitioning issue in the 
Everglades-to-bay freshwater delivery rather than errors in the 
marine boundary forcing, and is a target for future calibration 
refinement. A more thorough study on the spatial differences
in evaporation across basins, particularly as a function of
water temperature, may lead to further improvements. Sufficient 
evaporation data is not yet available for a detailed assessment.

Despite the bias, the model captures interannual variability in salinity 
across the domain with correlations ranging from $r = 0.373$ (Butternut
Key) to $r = 0.788$ (Rabbit Key). Basins with the strongest performance 
are those with direct and well-characterized connections to either
Everglades freshwater or tidal forcing. Long Sound ($r = 0.744$,
RMSE = 7.0 ppt) and Rankin 
Lake ($r = 0.669$, RMSE = 7.6 ppt) are highly freshwater driven and
at least partially isolated from broader marine influences and perform
well. The detail comparison at Long Sound 
(Figure~\ref{fig:salinity_detail}) shows that BAM correctly captures the 
direction and approximate timing of the seasonal freshwater signal, with 
summer salinity minima during wet seasons and recovery during dry seasons, 
consistent with the dominant Everglades forcing at that location.
Interestingly, performance along the marine margin can also be quite strong
with Rabbit Key ($r = 0.788$, RMSE = 2.6 ppt), driven by strong tidal exchange
with the Gulf, showing more accurate model results.

Joe Bay is a notable exception with respect to model performance, showing 
a persistent positive bias of $+$11.3  ppt and RMSE of 14.1 ppt  
throughout the 26-year record. This bias is present across all periods 
including pre- and post-instrument transition, confirming it as a 
calibration limitation rather than a data artifact. The Joe Bay shoreward 
shoal connection to the Everglades may be insufficiently transmissive in the 
current calibration, resulting in the model basin retaining a more marine 
character than the observations support. Recalibration of the Joe Bay 
shoreward boundary shoals is recommended as a priority improvement for 
future model versions.

For restoration planning purposes the model's ability to capture 
interannual variability in salinity response to changes in Everglades 
water levels is more relevant than absolute bias. The well-mixed basin 
assumption means that modeled salinity represents a spatial and temporal 
average while observations are point measurements, and some discrepancy 
is inherent in this comparison. The consistent interannual signal across 
the domain, combined with the near-zero domain-wide mean bias, supports 
the use of BAM for scenario-based assessment of the direction and relative 
magnitude of salinity responses to altered freshwater delivery.

\begin{table}[htbp]
  \caption{BAM salinity simulation skill metrics by basin over the full period
  of record 1999--2026. Metrics are computed against observed days only; $N$ is
  the number of days with a valid observation at the paired gauge. Bias and RMSE
  are in ppt. Basins sharing an observational gauge are indicated by identical
  gauge codes. $^a$Joe Bay shows a persistent positive bias throughout the
  record attributed to insufficient freshwater delivery through the shoreward
  shoal calibration at that basin; see text for discussion.}
  \label{tab:sal_skill}
  \centering
  \begin{tabular}{llrrrr}
    \toprule
        Basin & Gauge & Bias (ppt) & RMSE (ppt) & $r$ & $N$ \\
    \midrule
        \multicolumn{6}{l}{\textit{Northeastern Bay}} \\
    Long Sound                & LS & $+$3.53 & 6.96 & 0.744 & 9,606 \\
    Barnes Sound              & MD & $+$3.83 & 6.88 & 0.481 & 8,868 \\
    Manatee Bay               & MB & $+$1.98 & 5.44 & 0.611 & 8,869 \\
    Joe Bay$^a$               & JB & $+$11.31 & 14.12 & 0.663 & 9,018 \\
    Little Madeira Bay        & LM & $+$4.04 & 7.64 & 0.574 & 9,624 \\
    Blackwater Sound          & BS & $+$0.90 & 4.46 & 0.627 & 9,684 \\
    Little Blackwater Sound   & LB & $+$2.17 & 5.99 & 0.652 & 9,656 \\
    \midrule
    \multicolumn{6}{l}{\textit{Central Bay}} \\
    Butternut Key             & BN & $-$2.42 & 6.80 & 0.373 & 9,686 \\
    Deer Key                  & DK & $-$2.17 & 6.10 & 0.498 & 9,529 \\
    Duck Key                  & DK & $-$2.13 & 6.09 & 0.498 & 9,529 \\
    Lignumvitae               & PK & $-$1.77 & 3.24 & 0.674 & 9,534 \\
    Porpoise Lake             & BA & $-$2.19 & 5.15 & 0.528 & 9,624 \\
    Twin Keys                 & LR & $-$2.32 & 3.82 & 0.691 & 9,561 \\
    \midrule
    \multicolumn{6}{l}{\textit{Western Bay}} \\
    Whipray                   & WB & $-$5.87 & 8.40 & 0.517 & 9,707 \\
    Rankin Lake               & GB & $-$1.26 & 7.61 & 0.669 & 9,482 \\
    Johnson Key               & JK & $-$2.89 & 4.61 & 0.673 & 9,610 \\
    Snake Bight               & BK & $-$3.47 & 5.87 & 0.628 & 9,419 \\
    Rabbit Key                & LR & $-$1.27 & 2.64 & 0.788 & 9,561 \\
    \midrule
    \multicolumn{2}{l}{\textit{Domain mean}}
                              &    0.00 & 6.21 & 0.605 & \\
    \multicolumn{2}{l}{\textit{Domain mean absolute}}
                              &    3.08 &      &      & \\
    \bottomrule
  \end{tabular}
\end{table}

\subsection{Skill Over the Extension Period}

The calibration parameters -- per-shoal Manning's coefficients and the shoal
runoff geometry -- predate the period-of-record extension and were not revised
during it. Performance after 2017-01-01 is therefore a temporal hold-out, and
Table~\ref{tab:split_period} reports the two periods separately.

Skill in the hold-out period is indistinguishable from the calibration era.
Water level correlation falls by 0.004 and salinity correlation by 0.008, while
salinity RMSE improves from 6.53 to 5.54 ppt and the amplitude ratio is
unchanged to within 0.02. The model is performs well for a decade of
boundary conditions in the validation period, the property that matters
most for its intended use.

\begin{table}[htbp]
  \caption{BAM skill over the calibration era and the extension period, computed
  against observed days only. The 2017-01-01 split marks the end of the original
  period of record; calibration parameters were not revised during the
  extension, so the later period is an independent hold-out.
  $\sigma_m /   \sigma_o$ is the ratio of modelled to observed standard
  deviation, where values below one indicate damped variability. $N$ is the
  total number of scored basin-days across 15 water level and 18 salinity
  basins.}
  \label{tab:split_period}
  \centering
  \begin{tabular}{lrrrrr}
    \toprule
    Period & Bias & RMSE & $r$ & $\sigma_m / \sigma_o$ & $N$ \\
    \midrule
    \multicolumn{6}{l}{\textit{Water level (m)}} \\
    Calibration era, 1999--2016    & $-$0.028 & 0.090 & 0.786 & 1.10 & 91,446 \\
    Extension period, 2017--2026   & $-$0.017 & 0.093 & 0.782 & 1.10 & 50,178 \\
    \midrule
    \multicolumn{6}{l}{\textit{Salinity (ppt)}} \\
    Calibration era, 1999--2016    & $-$0.04 & 6.53 & 0.611 & 0.95 & 110,091 \\
    Extension period, 2017--2026   & $+$0.08 & 5.54 & 0.603 & 0.93 & 60,476 \\
    \bottomrule
  \end{tabular}
\end{table}

\subsection{Model Bias Across Hydrologic Regimes}

\begin{figure}[htbp]
  \centering
  \includegraphics[width=\textwidth]{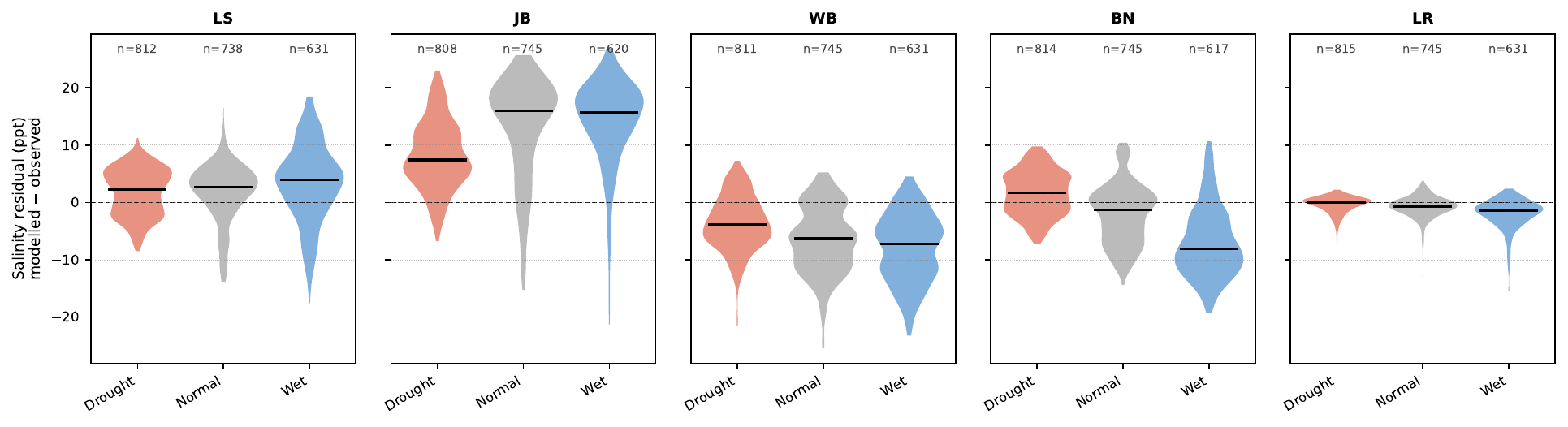}
  \caption{BAM salinity bias (modeled minus observed, ppt) at five
  representative basins partitioned by hydrologic regime over 2010--2015.
  Drought, normal, and wet regimes (red, gray, blue) are separated by fixed
  thresholds on domain-mean EDEN shoreward stage, taken as the \nth{33} and
  \nth{67} percentiles of the full period of record ($0.084$ m and $0.201$ m
  relative to Florida Bay mean sea level). Because the thresholds derive from
  the full record and this window is a subset of it, the three groups are
  unequal in size ($n = 815$, $745$, and $631$ days). Residuals are
  computed on observed days only. Horizontal bars indicate median bias.
  Bias magnitude increases from drought toward wet conditions at Joe Bay (JB)
  and Whipray (WB), and reverses sign at Butternut Key (BN).}
  \label{fig:drought_wet}
\end{figure}

A drought and wet season bias analysis provides additional context for
interpreting model performance under the range of hydrologic conditions
that characterize the 2010--2015 evaluation period. EDEN stage values
were partitioned into drought, normal, and wet periods and model bias
(modeled minus observed salinity) was examined separately for each
regime at the five representative basins (Figure~\ref{fig:drought_wet}).

Bias magnitude varies systematically with hydrologic regime rather than
remaining constant. Median bias at Joe Bay (JB) rises from $+$7.4 ppt under
drought conditions to $+$16.0 ppt under normal and $+$15.7 ppt under wet
conditions, and Butternut Key (BN) reverses sign entirely, from $+$1.7 ppt
under drought to $-$8.1 ppt under wet. Whipray (WB) becomes progressively
more negative, from $-$3.9 to $-$7.3 ppt, while Long Sound (LS) and Rabbit
Key (LR) vary by less than 1.6 ppt across all three regimes. The direction 
of this drift is consistent with the spatial bias structure of
Section~\ref{sec:salinity_performance}: the model delivers too little
freshwater to the northeastern margin and too much marine water to the central
and western bay, and both errors grow as more water is routed through the
domain. Anomaly-based (scenario minus baseline) metrics therefore remain the
appropriate basis for scenario comparison, but the assumption that model error
cancels between a scenario and its baseline holds only where the two share a
similar hydrologic regime.

\FloatBarrier

\section{Illustrative Application}
\label{sec:scenarios}

The purpose of this section is to demonstrate that the extended period of
record is usable for scenario work, not to analyze any particular restoration
question. A single example is presented: a uniform offset applied to the
Everglades shoreward boundary, which exercises the linkage between upstream
water levels and bay salinity that the extension was built to support. A fuller
treatment of restoration alternatives and sea level rise, including their
interaction, is reported separately \citep{Stabenau2026ESCO}. While the upstream
stage could equally be driven by output from the RSM, the SFWMM, or another
marsh model, that option is not exercised here.

\subsection{Baseline Simulation}

The baseline simulation spans 2010-01-01 to 2015-12-31, a six-year
window that captures a range of hydrologic conditions including the
2011 drought, the 2012--2013 wet period, and the 2015 El~Ni\~{n}o
signal. All boundary conditions are drawn from observed data: EDEN
stage at the eight shoreward boundary stations, tidal harmonics and
observed MSL anomaly at the marine margins, and observed rainfall and
ET at the basin level. The baseline provides the reference salinity
timeseries against which the scenario anomalies of
Section~\ref{sec:inflow_sensitivity} are computed.

\subsection{Sensitivity to Freshwater Inflow}
\label{sec:inflow_sensitivity}

Freshwater delivery from the Everglades to the coastal basins of Florida Bay is
the primary management lever available through CERP. BAM provides a direct means
of translating proposed changes in Everglades water levels into projected
salinity responses across the 54-basin domain. To illustrate this sensitivity, a
series of scenarios was constructed by applying uniform stage offsets to the
EDEN shoreward boundary conditions, representing a simplified analog to the
water level changes that might result from upstream restoration actions. Stage
offsets of $-$0.15 m (FW-Low), $+$0.15 m (FW-Mid), and $+$0.30 m (FW-High) were
applied uniformly across all eight EDEN boundary stations relative to the
baseline simulation, with all other boundary conditions held constant.

\begin{figure}[htbp]
  \centering
  \includegraphics[width=\textwidth]{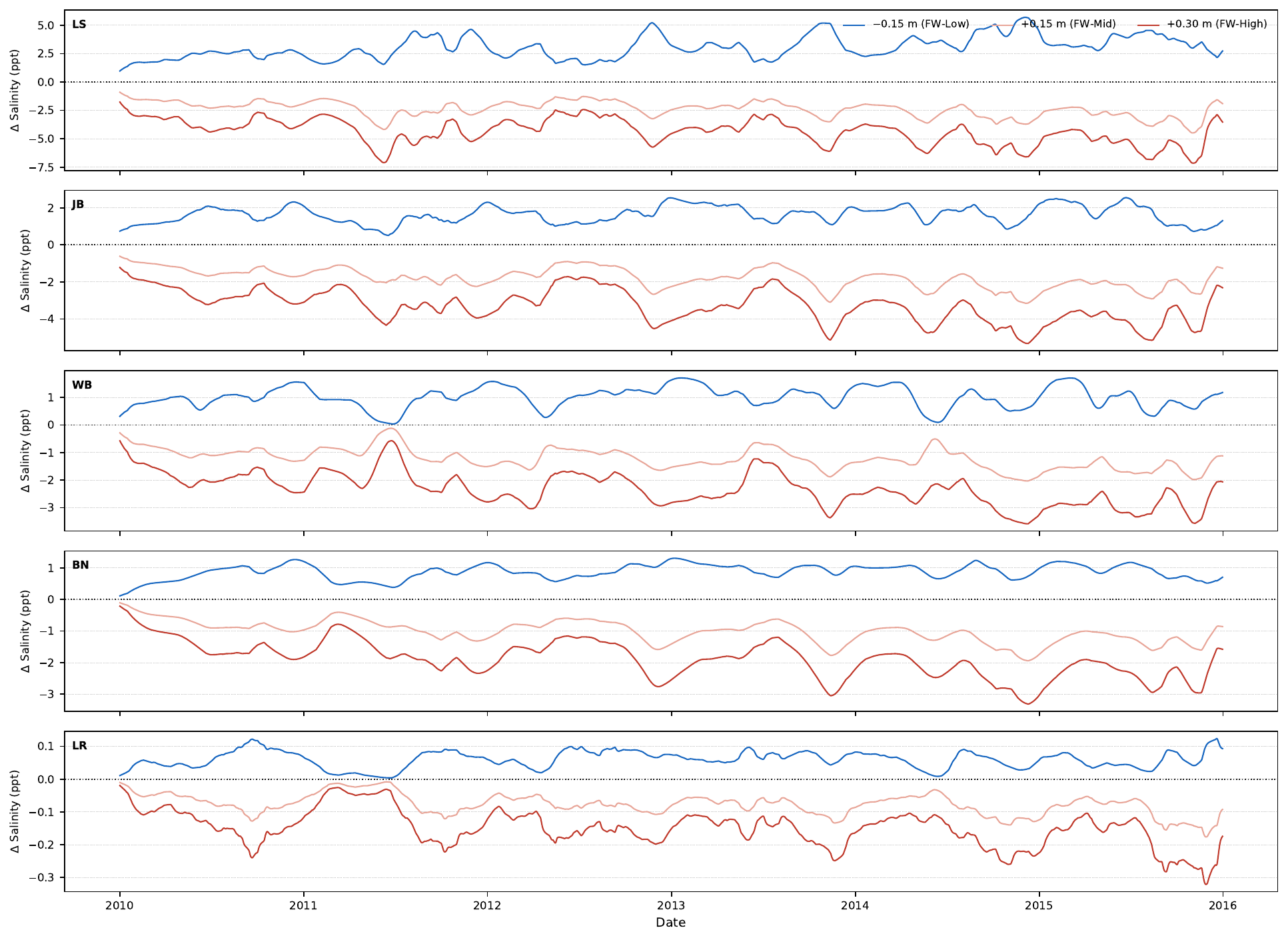}
  \caption{BAM salinity anomaly (scenario minus baseline) at five representative
  basins under prescribed uniform offsets to Everglades shoreward stage:
  $-$0.15 m (FW-Low, blue), $+$0.15 m (FW-Mid, light red), and $+$0.30 m
  (FW-High, dark red), evaluated over 2010--2015. Northeastern bay basins
  (LS, JB) show the largest and most direct response to altered freshwater
  delivery. Central bay basins (WB, BN) show attenuated responses. The
  western marine basin (LR) is largely insensitive to shoreward stage changes.
  Seasonal modulation is evident in all freshwater-influenced basins.
  Traces are 30-day centred rolling means. Quoted statistics are computed on
  the unsmoothed daily anomaly.}
  \label{fig:inflow_scenario}
\end{figure}

The salinity response to altered freshwater inflow is not uniform across the bay
(Figure~\ref{fig:inflow_scenario}). Basins along the northeastern margin show
the strongest response: Long Sound (LS) exhibits mean anomalies of $+$3.1 ppt
under FW-Low, $-$2.4 ppt under FW-Mid and $-$4.4 ppt under FW-High, while Joe
Bay (JB) shows a similarly scaled but attenuated response of $+$1.6 and $-$3.3
ppt respectively. Central bay basins (WB, BN) show moderate responses of $-$1.9
to $-$2.3 ppt under FW-High, while the western marine basin Rabbit Key (LR) is
nearly insensitive to shoreward stage changes, with mean anomalies within
$\pm$0.15 ppt and extremes within 0.6 ppt across all scenarios. This spatial
gradient reflects the progressive attenuation of freshwater signals through the
interbasin shoal network with increasing distance from the shoreward boundary.

Salinity anomalies scale sublinearly with the magnitude of the stage offset, a
doubling of the offset producing 1.8--1.9 times the response, and exhibit clear
seasonal modulation, with the largest anomalies
occurring during the wet season (June--October) when freshwater gradients
across the shoreward boundary are strongest. These results confirm that BAM
captures the expected physical response to altered freshwater delivery and
support its use for scenario-based assessment of restoration outcomes.

\section{Discussion}
\label{sec:discussion}

BAM is deliberately simple. The model's 54-basin decomposition, single-layer
hydraulics, and instantaneous mixing assumption represent a conscious trade-off
between physical fidelity and operational utility. As Box famously noted, all
models are wrong, but some are useful \citep{Box1976}. BAM's utility lies not in
its capacity to resolve the detailed hydrodynamics of Florida Bay, but in its
ability to translate changes in upstream Everglades water levels into
first-order salinity projections across the bay with minimal computational
overhead and complete transparency of implementation. The 26-year performance
record presented in Section~\ref{sec:performance} supports this claim: despite
its simplicity, the model captures the dominant interannual and seasonal
salinity signals across the domain with sufficient fidelity for restoration
scenario assessment.

\subsection*{Model Performance in Context}

The domain-wide salinity skill ($r = 0.605$, RMSE = 6.21 ppt, bias = 0.00 ppt)
is best interpreted relative to the model's intended application rather than
against the standards of a high-resolution hydrodynamic model. For restoration
planning, the relevant question is whether the model correctly predicts the
\textit{direction} and \textit{relative magnitude} of salinity change in
response to altered freshwater delivery rather than reproducing the absolute
salinity at any point on any given day. The near-zero domain-wide mean bias,
combined with correlations ranging up to $r = 0.788$ in the best-resolved
basins, indicates that BAM meets this standard.

The opposing spatial bias structure, which is positive in the northeastern bay
and negative in the central and western bay, is the most significant systematic
limitation revealed by the performance evaluation. This pattern is consistent
with the distribution of freshwater delivery across the shoreward boundary
rather than errors in marine forcing, and points to the shoal connectivity along
the northeastern margin as the primary target for future recalibration. The Joe
Bay bias ($+$11.3 ppt) is an isolated and well-characterized exception that
likely reflects insufficient transmissivity in the shoreward shoal at that
basin. It does not propagate meaningfully into adjacent basins and its influence
on domain-wide scenario projections is notable but limited.

Water level performance ($r = 0.797$, RMSE = 0.092 m across 15 gauged basins) is
adequate for driving mixing and therefore calculation of salinity based on
transport across the domain. The near-zero mean bias confirms that the datum
conversion framework described in Section~\ref{sec:stage_performance}
appropriately aligns modeled and observed water levels, which is a prerequisite
for using upstream model output (RSM, SFWMM) as a direct substitute for observed
data at the shoreward boundary.

\subsection*{Comparison to Prior Modeling Approaches}

BAM occupies a distinct position in the landscape of Florida Bay 
salinity modeling tools. Marshall et al. \citeyearpar{Marshall2008} 
reviewed the suite of models available for Florida Bay salinity 
simulation prior to BAM's development, ranging from statistical 
regression approaches to full three-dimensional hydrodynamic models. 
Empirical tools, such as those described in 
\citet{Marshall2011}, provide efficient salinity estimates from 
stage and flow data but lack the physical process representation 
needed to project salinity responses to novel boundary conditions 
outside the range of the calibration record. Full hydrodynamic 
models provide higher spatial and temporal resolution but require 
substantially greater computational resources, expertise, and data 
inputs, limiting their accessibility for routine restoration planning.
To date, operationally accessible hydrodynamic models for 
Florida Bay salinity assessment have not been documented in the 
literature \citep{Marshall2008}.

BAM's immediate predecessor, FATHOM \citep{Cosby2010}, shares the 
same basin-and-shoal domain representation and Manning's hydraulics. 
BAM improves on FATHOM in several important respects: it replaces 
FATHOM's compiled Fortran codebase with a fully open-source Python 
implementation; it incorporates a substantially extended and 
higher-quality observational record for both calibration and 
boundary forcing; and it provides a direct, documented pathway for 
substituting upstream restoration model output at the shoreward 
boundary. These improvements lower the barrier to use for restoration 
planners and increase the model's transparency and reproducibility.

\subsection*{The Python Implementation}

The decision to implement BAM entirely in Python has practical 
consequences beyond mere convenience. A compiled legacy codebase 
presents a significant barrier to community engagement: modifying 
parameters, inspecting intermediate calculations, or adapting the 
model for new applications requires either access to the original 
source and compiler toolchain, or acceptance of the model as an 
opaque binary. BAM's Python implementation eliminates these barriers. 
Every parameter, equation, and output is accessible to any user with 
a standard Python environment, with no licensing costs and no 
platform dependencies. This transparency is particularly valuable in 
the context of regulatory and planning applications, where the ability 
to inspect and verify model behavior is often as important as the 
model results themselves.

The primary operational cost of the Python implementation is runtime. 
Full period-of-record simulations (26 years at 6-minute timesteps) 
require several hours on a standard workstation. For scenario 
assessment applications, where multiple runs over shorter evaluation 
windows are more typical, this is not a meaningful constraint.

\subsection*{Appropriate and Inappropriate Uses}

BAM is well suited for evaluating the direction and relative 
magnitude of salinity responses to changes in Everglades freshwater 
delivery and mean sea level. It is the appropriate tool when the 
question is: given a proposed change to upstream water management, 
how will salinity in Florida Bay respond, and which basins will be 
most affected? The model is less appropriate, and should not be used, 
for applications requiring accurate simulation of storm surge, wave 
propagation, sub-basin spatial salinity gradients, or rapid 
transient responses at timescales shorter than tidal. The well-mixed 
basin assumption and the absence of vertical structure also preclude 
meaningful simulation of stratification or density-driven circulation. 
Within its intended scope, however, BAM provides a reliable, 
transparent, and computationally efficient tool for the restoration 
planning community.

\section{Conclusions}
\label{sec:conclusions}

The Bay Assessment Model provides a transparent, observation-driven
salinity projection tool for Florida Bay, calibrated against a
26-year record of hydrographic observations from the Everglades National Park
Marine Monitoring Network. The principal findings of this overview are as
follows. Model performance over the full 1999--2026 period of record is adequate
for scenario-based restoration assessment. Domain-wide mean salinity bias is
0.00 ppt with a mean RMSE of 6.21 ppt and mean Pearson correlation of
$r = 0.605$ across 18 basins. Water level simulation achieves a domain-wide mean
correlation across 15 gauged basins of $r = 0.797$ and RMSE of 0.092 m. The
model effectively captures the dominant seasonal and interannual salinity signals
across the domain, including the freshwater-driven salinity minima in the
northeastern bay and the more marine-dominated regime of the western and central
basins.

The illustrative application demonstrates that BAM responds
physically consistently to prescribed changes in Everglades
freshwater delivery: uniform stage increases of 0.15 and 0.30 m at
the shoreward boundary produce salinity reductions of 2--4 ppt in the
northeastern bay with progressively attenuated responses toward the
marine margins. Model bias grows in magnitude from drought toward wet
conditions and reverses sign at one central bay basin, so anomaly-based
comparison is sound only between runs sharing a similar hydrologic regime.
Restoration and sea level scenarios are treated in detail elsewhere
\citep{Stabenau2026ESCO}; the contribution here is the extended and
documented record on which such analyses depend.

The Python implementation and open-source availability of BAM lower
the barrier to use for restoration planners, enable community
development, and ensure complete transparency of model behavior.
BAM's direct linkage between EDEN stage observations and Florida Bay
salinity means that output from upstream restoration planning models
such as RSM or SFWMM can be substituted at the shoreward boundary
without modification, making BAM a natural and efficient downstream
component of the CERP scenario assessment workflow.

The BAM source code, input data, and operational manual remain
freely available, as detailed in Appendix A.

\section*{Acknowledgments}

Water level, salinity, rainfall and temperature observations were
collected and maintained by the Marine Monitoring Network of Everglades
National Park. Everglades water levels were obtained from the Everglades
Depth Estimation Network, maintained by the U.S. Geological Survey.
Tidal harmonic constituents and sea level records were obtained from the
NOAA Center for Operational Oceanographic Products and Services, and
structure flow records from the South Florida Water Management District.
BAM was originally developed and released by the Software Literacy
Foundation.

Portions of this effort were performed using Claude Code (Anthropic), which was
used to develop and benchmark the gap-filling procedure described in
Section~\ref{sec:mmn}, to compute the skill metrics reported in
Section~\ref{sec:performance}, and to generate the figures. All methods,
results and interpretations were reviewed and verified by the authors, who take
full responsibility for the content of this paper.

\bibliographystyle{apalike}  
\bibliography{references}

\appendix
\section*{Appendix A: Model Availability}

BAM is freely available as an open-source project licensed under the 
GNU General Public License v3 (GPL v3) at:
\begin{center}
\url{https://github.com/SoftwareLiteracyFoundation/BAM}
\end{center}
The repository includes the complete Python source code, example 
input data, and configuration files. Installation and command-line 
usage are described in detail in the operational manual 
\citep{Park2016}. Users wishing to apply BAM to new domains or 
substitute upstream model output at the shoreward boundary should
consult the manual for guidance on input file formats and
command-line options. The input files used for the simulations reported here
are listed in Table~\ref{tab:inputs}.

\end{document}